\documentclass[a4paper,fleqn,usenatbib]{mnras}

\usepackage{newtxtext,newtxmath}

\usepackage{bm}
\usepackage{booktabs}
\usepackage{threeparttable}

\usepackage{setspace}

\usepackage[T1]{fontenc}
\usepackage{ae,aecompl}

\usepackage{graphicx}	
\usepackage{amsmath}	
\usepackage{amssymb}	
\usepackage{xcolor}

\definecolor{pink}{RGB}{255,20,147}
\definecolor{purple}{RGB}{160,32,240}
\definecolor{gold}{RGB}{0,139,0}
\usepackage{times}

\usepackage{amsfonts}

\newcommand{\angstrom}{\mbox{\normalfont\AA}}

\title[Magnetic-reconnection-heated corona in AGNs]{Magnetic-reconnection-heated corona in active galactic nuclei: refined disc--corona model and application to broad-band radiation}
\author[H. Cheng et al.]{
Huaqing Cheng,$^{1,2}$\thanks{E-mail: hqcheng@nao.cas.cn}
B. F. Liu,$^{1,2}$\thanks{E-mail: bfliu@nao.cas.cn}
Jieying Liu,$^{3,4,5}$
Zhu Liu,$^{1}$
Erlin Qiao,$^{1,2}$
\newauthor
and Weimin Yuan$^{1,2}$
\\
$^{1}$Key Laboratory of Space Astronomy and Technology, National Astronomical Observatories, Chinese Academy of Sciences, Beijing 100012, P. R. China\\
$^{2}$School of Astronomy and Space Sciences, University of Chinese Academy of Sciences, Beijing 100049, P. R. China\\
$^{3}$National Astronomical Observatories/Yunnan Observatory, Chinese Academy of Sciences, Kunming 650011, P. R. China\\
$^{4}$Key Laboratory for the Structure and Evolution of Celestial Objects, Chinese Academy of Sciences, Kunming 650011, P. R. China\\
$^{5}$Center for Astronomical Mega-Science, Chinese Academy of Sciences,  Beijing 100012, P. R. China}
\date{Accepted XXX. Received YYY; in original form ZZZ}

\pubyear{2019}

\begin{document}
\label{firstpage}
\pagerange{\pageref{firstpage}--\pageref{lastpage}}
\maketitle

\begin{abstract}

A long-standing question in active galactic nucleus (AGN) research is how the corona is heated up to produce X-ray radiation much stronger than that arising from the viscous heating within the corona. 
In this paper, we carry out detailed investigations of magnetic-reconnection heating to the corona, specifically, studying
how the disc and corona are self-consistently coupled with the magnetic field, and how the emergent spectra depend on the fundamental parameters of AGN. 
It is shown that diverse spectral shapes and luminosities over a broad bandpass from optical to X-ray  
can be produced from the coupled disc and corona within a limited range of the black hole mass, accretion rate and magnetic field strength. The relative strength of X-ray emission with respect to optical/ultraviolet (UV) depends on the strength of the magnetic field in the disc, which, together with accretion rate, determines the fraction of accretion energy transported and released in the corona. This refined disc--corona model is then applied to reproduce the broad-band spectral energy distributions (SEDs) of a sample of 20 bright local AGNs observed simultaneously in X-ray and optical/UV. 
We find that, in general, the overall observed broad-band SEDs can be reasonably reproduced,  except for rather hard X-ray spectral shapes in some objects.
The radiation pressure-dominant region, as previously predicted for the standard accretion disc in AGN, disappears for strong X-ray sources,
revealing that AGN accretion discs are indeed commonly stable as observed. Our study suggests the disc--corona coupling model involving magnetic fields to be a promising approach for understanding the broad-band spectra of bright AGNs. 

\end{abstract}





\section{Introduction}

It is widely believed that the radiation of the active galactic nuclei (AGNs) is powered by the accretion of matter onto a supermassive black hole. The broad-band spectral energy distributions (SEDs) in high-luminosity AGNs are usually interpreted as radiation from a geometrically thin, optically thick accretion disc sandwiched in a geometrically thick, optically thin corona. Specifically,  the optical/UV spectra, which are often referred to as the `big blue bump', are contributed by the thermal emissions from the cold disc \citep[e.g.][]{1978Natur.272..706S, 1983ApJ...268..582M}, while the power-law emissions in hard X-ray band are produced by thermal electrons in the hot corona  through Comptonization of soft photons from the disc \citep[e.g.][]{1994ApJ...436..599S, 1998MNRAS.301..179M}. 

Interaction between the cold disc and the hot corona is essential. Theoretical models elucidating the interactions have been studied for decades \citep[e.g.][]{1991ApJ...380L..51H, 1993ApJ...413..507H, 1993PASJ...45..775N, 1994ApJ...436..599S, 1996ApJ...470..249P, 1997ApJ...487..747D, 2002ApJ...572L.173L, 2003ApJ...587..571L, 2004ApJ...607L.107W, 2009MNRAS.394..207C, 2012ApJ...761..109Y, 2016RAA....16...55Y, 2016ApJ...833...35L, 2018A&A...614A..79P}. Of these investigations the radiation coupling between the disc and corona, that is the Comptonization of disc photons in the corona and the coronal irradiation to the disc, has been well considered. Heat conduction and enthalpy/mass flow between the hot corona and cold disc  were also taken into account in some of the studies \citep[e.g.][]{2000A&A...361..175M, 2002ApJ...575..117L}. Detailed studies reveal that the corona is weaker at higher accretion rates because efficient Compton cooling leads to gas condensation to disc, predicting weak X-ray emission at high states \citep[e.g.][]{2002ApJ...575..117L,2012A&A...544A..87M}.   Even if the condensation of hot gas is neglected,  viscous heating within the corona is insufficient to produce  the observed X-ray luminosity. In order to account for the strong X-ray radiation in luminous AGNs, it is often assumed that the energy released by disc accretion is somehow transported to the corona and radiates in X-rays. Hence, additional heating mechanism has been explored to keep the corona sufficiently strong to produce the observed X-ray luminosity.

One of the most promising mechanisms for the coronal heating is through magnetic field \citep[e.g.][]{1992MNRAS.259..604T, 1998MNRAS.299L..15D, 2000ApJ...534..398M, 2001MNRAS.321..549M, 2002ApJ...572L.173L, 2003ApJ...587..571L, 2003MNRAS.341.1051M, 2004ApJ...607L.107W, 2009MNRAS.394..207C, 2012ApJ...761..109Y}. The concept of magnetic-reconnection heating is initiated from solar flares \citep[e.g.][]{1999ApJ...526L..49S}. The magnetic field is generated by the dynamo process in the disc and emerges into the corona as a result of Parker buoyancy instability.  The stored magnetic energy is then released in the corona via magnetic-reconnection. In this manner, part of the accretion energy liberated in the disc is transferred to the corona through magnetic field, heating up the corona during reconnection and eventually radiated away in X-rays.

A self-consistent disc--corona model was proposed in the frame of magnetic-reconnection heating and disc evaporation \citep{2002ApJ...572L.173L, 2003ApJ...587..571L}. In this model, the corona is assumed to be heated by the aforementioned magnetic-reconnection, which is balanced by inverse Compton (IC) scattering. Heat conducted from the hot corona to the disc surface leads to continuous evaporation, replenishing gas to the corona during its accretion towards the black hole. Specifically, the fraction of the energy liberated in the corona, $f$, is physically determined  by the coupling of the disc and corona involving magnetic fields, which was usually taken as a free parameter in previous studies. In addition, the disc evaporation provides a natural and reasonable explanation for the mass supply of the hot corona. Therefore, such a model provides a physical mechanism for both the formation and heating of the corona. Numerical calculations show that a significant fraction of accretion energy can be transported to the corona, making it possible for the corona to produce strong X-ray emissions as observed \citep{2003ApJ...587..571L, 2015MNRAS.448.1099Q}. The model has been further developed to more general cases, and is expected to be applicable to the variety of observed spectra in AGNs \citep{2016ApJ...833...35L}. Nevertheless, these earlier studies mainly focused on X-ray emission originating from a small region near the black hole, neglecting the contribution of outer disc emission to the optical/UV band.  Radiation from the entire accretion flow needs to be taken into account when the overall spectrum is modelled. Also, the self-consistency of the model needs to be improved to a higher precision. These refinements are crucial for the model to be realistically applicable to modelling observational data, and are one of the motivations of this work.

On the other hand, we have recently conducted a study of the broad-band SEDs for a well-selected sample of $23$ Seyfert 1 galaxies, by making use of their simultaneous optical, UV and X-ray observations with the Neil Gehrels {\it Swift} Observatory \citep{2019MNRAS.487.3884C}. The physical properties of the objects in the sample are found to spread over a broad range, for example the black hole mass $M_{\rm BH}\approx10^7$--$10^9M_{\sun}$ and Eddington ratio $\lambda_{\rm Edd} \equiv L_{\rm bol}/L_{\rm Edd} \approx0.01$--$1$, which provides a reasonable pre-requisite for the application of the disc--corona model. 

The motivation of this work is to understand quantitatively the physical origin of the broad-band SEDs of AGN, in the framework of the magnetic-reconnection-heated corona model \citep{2002ApJ...572L.173L, 2003ApJ...587..571L, 2016ApJ...833...35L}. First,  the model is further improved and refined (see Section \ref{sec:disccorona}) to be more realistic and applicable to compare directly with observations. Then, the model is applied to reproducing the observed broad-band SED of the aforementioned AGN sample. 
We find that for the majority of the sample objects, the observed 
SEDs can 
generally be reproduced by the disc coupled with a strong magnetic 
reconnection-heated corona,
though for some the X-ray spectra are too flat to be accounted for. 

The paper is organized as follows. In Section \ref{sec:disccorona}, the magnetic-reconnection-heated corona model is briefly described and illustrative calculations of the spectrum for typical AGNs are presented. In Section \ref{sec:modelsed}, the spectral modelling of the broad-band SEDs  is described and main results for individual sources are shown and illustrated. The discussion and conclusions are given in Section \ref{sec:discuss} and \ref{sec:sum}.

\section{magnetic-reconnection-heated corona}\label{sec:disccorona}
\subsection{Model descriptions} \label{sec:model}

A magnetic-reconnection-heated corona model is proposed in \citet{2002ApJ...572L.173L, 2003ApJ...587..571L} and revisited by \citet{2016ApJ...833...35L}.  In this model, it is assumed that a standard  geometrically thin and optically thick disc \citep{1973A&A....24..337S} extends to the innermost stable circular orbit,  sandwiched by a hot geometrically thick and optically thin, accreting corona. The magnetic field is assumed to be generated by the dynamo process operating in the accretion disc. Under the influence of Parker instability, the magnetic loops emerge into the corona and reconnect with other loops. In this manner, some fraction ($f$) of the accretion energy stored in the magnetic field is transferred into the corona and converted into thermal energy of the electrons. This energy is eventually radiated through the process of IC scattering. The disc is heated partially by accretion and partially by illumination from the corona.  In the chromospheric layer between the disc and corona, thermal conduction heats up gas, leading to the disc gas evaporating into the corona and supplying for the steady accretion of coronal flow \citep[e.g.][]{2000A&A...354L..67M, 2000A&A...361..175M,2000A&A...360.1170R, 2002A&A...387..918S}. The disc and corona are coupled through the magnetic field, the gas evaporation and coronal illumination to the disc. The structure of disc and corona can be self-consistently determined by taking into account this coupling, from which the emergent spectrum can be calculated by Monte Carlo simulations. 

Note that the magnetic heating to coronal electrons is much more efficient than heating through Coulomb collisions with ions,  the latter is neglected in our calculation of electron temperature.  The ions in the corona are heated up to a higher temperature by viscously released energy  as they accrete to the black hole, leading to a two-temperature corona.  Such an approximation  is  justified by our calculations for  bright AGN, and is, in principle,  valid  for sources with X-ray emission stronger than that from a pure advection-dominated accretion flow \citep[ADAF, ][]{1994ApJ...428L..13N}.

The equations determining the structure of the accretion flows have been list in detail in \citet{2016ApJ...833...35L},  which are summarized as follows.  

Equations determining  the coronal density and electron temperature  are, 
\begin{eqnarray}
\label{eq:coronaenergy}
\frac{B^2}{4\upi}V_{\rm A}\approx \frac{4kT_{\rm e}}{m_{\rm e}c^2}\tau cU_{\rm rad},
\end{eqnarray}

\begin{eqnarray}\label{eq:evap}
{k_0T_{\rm e} ^{7\over 2}\over \ell_{\rm c}}\approx
{\gamma\over \gamma-1} n_{\rm c} k T_{\rm e}
\left(\frac{kT_{\rm e}} {\mu m_{\rm H}}\right)^{1/2}.
\end{eqnarray}
Eq.(\ref{eq:coronaenergy}), describes the energy balance in the corona. The magnetic loops emerge at Alfv\'{e}n speed $V_{\rm A}\equiv\sqrt{B^2/4\upi\mu m_{\rm H}n_{\rm c}}$, bringing magnetic energy flux $\frac{B^2}{4\upi}V_{\rm A}$ into corona and being radiated by IC scattering.   Here $T_{\rm e}$,  $n_{\rm c}$, and $\tau$ are the coronal temperature, density, and modified scattering depth, respectively;  $U_{\rm rad}$ is the energy density of soft photons.
Eq.(\ref{eq:evap}) describes the energy balance of conduction-induced evaporation, where $\ell_{\rm c}$ is the length of magnetic loops in the corona.
Eqs.(\ref{eq:coronaenergy}) and (\ref{eq:evap}) determine the coronal density and electron temperature for given  soft photon field and magnetic field, which are coupled with the thin disc. 

Constants in above equations are  the Boltzmann constant $k=1.38\times10^{-16} \rm erg~K^{-1}$,  the thermal conduction coefficient $k_0=10^{-6}\ \rm erg~cm^{-1}~s^{-1}~K^{-7/2}$,  the ratio of specific heats $\gamma=5/3$,  the molecular weight for pure hydrogen plasma $\mu=0.5$,  the mass of hydrogen atom $m_{\rm H}$, the mass of electron $m_{\rm e}$ and the light speed $c$. 
The length of magnetic loops is set to be the vertical scale height of corona, $\ell_{\rm c}\approx H_{\rm c}$, as the loops emerge from the disc  and then expand  in the corona, where  $H_{\rm c}\approx R$ for an optically thin, two-temperature hot accretion flow. The scattering depth is modified by the average optical depth for the
isotropic incident photons undergoing upscattering in a sandwich corona geometry and the multiple scattering of soft photons, $\tau=\lambda_{\rm t}n_{\rm c}\sigma_{\rm T} H_{\rm c}$, where  $\lambda_{\rm t}$ is  larger than 1.

The energy density of soft photons ($U_{\rm rad}$) and the magnetic field ($B$) in Eq.(\ref{eq:coronaenergy}) are determined by the coupling between the disc and corona. The soft photons are contributed by disc accretion and coronal illumination,

\begin{eqnarray}
\label{eq:disc}
U_{\rm rad}={2\over c}\left\{\frac{3GM\dot{M}}{8\upi R^3}\left[1-\left(\frac{3R_{\rm S}}{R}\right)^{1/2}\right]-\frac{B^2}{4\upi}V_{\rm A}\right\}
+U_{\rm re},
\end{eqnarray}
where the first term on the right-hand side of the equation denotes the density of net energy from accretion gain and magnetic loss; The second term, $U_{\rm re}$, denotes energy density originating from local corona illumination.  When the hard photons from the corona have been reprocessed in the disc and emit into the corona, the energy density of these soft photons is proportional to the energy density of corona emission, with a coefficient relevant to the fraction of downward propagation and  reflection albedo ($\propto {1\over2} (1-a)$ given that the IC scattering is isotropic). Since the coronal radiation energy is eventually from magnetic energy,   $U_{\rm re}$ can be approximated as   $U_{\rm re}=0.4\lambda_{\rm u}\frac{B^2}{8\upi}$, where $a=0.2$ is adopted, and $\lambda_{\rm u}$ includes the deviation from isotropic scattering and the ratio of the speed of magnetic energy release (at  Alfv\'{e}n speed)  and radiation (at light speed).  The value of $\lambda_{\rm u}$ is around unit in order of magnitude, which is determined from Monte Carlo simulation in our calculations. 

The strength of magnetic field is  parametrized with a magnetic equipartition coefficient $\beta_0$, which is defined as the ratio of the
gas plus radiation pressure to the magnetic pressure ($P_{\rm B}={B^2\over 8\upi}$), 
\begin{eqnarray}\label{eq:equip}
{B^2\over 8\upi}={1\over \beta_0}\left({\frac {\rho_{\rm d} kT_{\rm d}}{\mu
m_{\rm p}}}+{\frac {1}{3}}aT_{\rm d}^4\right), \label{gas state}
\end{eqnarray}
therefore, the magnetic field can be determined once the disc density and mid-plane temperature ($\rho_{\rm d},T_{\rm d}$) are derived from the disc solutions for given mass of black hole, accretion rate, and viscosity.

It should be noted that a fraction ($f$) of the accretion energy is taken away from the disc, 
\begin{eqnarray}
f\equiv{\frac{B^2}{4\upi}V_{\rm A} \over {3GM\dot M \over 8\upi R^3}\left[{1-\left({3R_{\rm S}\over R}\right)^{1/2}}\right]},
\end{eqnarray}
where $f$ is an implicit function of $M,\dot M, \alpha, \beta_0$ as from Eq.(\ref{eq:equip}). This leads to corresponding change in the equations of energy conservation and angular momentum conservation for a steady accretion flow, that is,  
\begin{eqnarray}\label{eq:energy-d}
{3GM\dot M (1-f)\over 8\upi R^3}\left[{1-\left({3R_{\rm S}\over R}\right)^{1/2}}\right]
={4\sigma T_{\rm d}^4\over 3(\kappa_{\rm es}+\kappa_{\rm ff})H_{\rm d}}, 
\end{eqnarray}

\begin{eqnarray}\label{eq:angular-d}
\dot{M}(1-f)\Omega\left[1-\left({\frac {3R_{\rm S}}{R}}\right)^{1/2}\right]=4\upi H_{\rm d}\alpha P_{\rm d},
\end{eqnarray}
where the coefficients for the electron scattering and absorption are $\kappa_{\rm es}=0.4\,\rm ~{cm^{2}~g^{-1}}$, $\kappa_{\rm ff}=6.4\times10^{22}\rho_{\rm d}T_{\rm d}^{-7/2}\,\rm ~{cm^{2}~g^{-1}}$; $H_{\rm d}$ is the disc thickness, $H_{\rm d}= c_{\rm s}/{\Omega}=\sqrt{P_{\rm d}/\rho_{\rm d}}/{\Omega}$, and $P_{\rm d}$ is the total pressure exerted by gas, radiation and magnetic field, which can be expressed as function of $T_{\rm d}, \rho_{\rm d}$. 

Eqs. (\ref{eq:coronaenergy})--(\ref{eq:angular-d}) are the complete set of equations for seven unknowns, that is the electron temperature $T_{\rm e}$ and density $n_{\rm c}$ in the corona, the mid-plane temperature $T_{\rm d}$ and density $\rho_{\rm d}$ in the disc, the magnetic field strength $B$,  the energy fraction released in the corona $f$ and the soft photon energy density $U_{\rm rad}$ emitted from the disc.
Given the black hole mass $M$, accretion rate $\dot{M}$, viscosity parameter $\alpha$ and magnetic  coefficient $\beta_{\rm 0}$, the equations from  (\ref{eq:coronaenergy}) to (\ref{eq:angular-d}) can be numerically solved for different radius with initial value for $\lambda_{\rm t}=1.0$ and $\lambda_{\rm u}=1.0$. 
With the obtained disc and corona parameters we calculate the emergent spectra of the disc--corona system by Monte Carlo simulations.  We further check the self-consistency from the Monte Carlo results, that is,  (1) whether the luminosity of downward-scattered photons is approximately equal to irradiation luminosity assumed
 in the structure calculation (relevant to $\lambda_{\rm u}$), and (2)
 whether the luminosity of  escaped photons, composed of non-scattered, upward-scattered, and reflected photons, is approximately equal to the liberated rate of total gravitational energy. We adjust the two parameters $\lambda_{\rm t}$ and $\lambda_{\rm u}$ in the iterative computation until the above two conditions are fulfilled. This iterative calculation modifies the density of re-processed soft photons and the scattering depth adopted in calculating the disc and  corona parameters by Monte Carlo simulations.  For a detailed explanation on the iterative procedure, see \citet{2003ApJ...587..571L}.
  
In this work, significant improvements on our disc--corona model have been made. First, the radiation from the the whole stable disc--corona system is  taken into account. The outer boundary of the disc--corona system is set to the self-gravity radius $R_{\rm sg}$ of several hundreds to thousand $R_{\rm S}$ \citep{1989MNRAS.238..897L}. Whilst in the previous work only the inner part of $R\le50R_{\rm S}$ was considered when the study focused on X-ray emission.  This is crucial for a more realistic description of the emergent spectra in the optical band, as this portion of the broad-band SED is mainly contributed by the thermal emissions at larger distances. Second, 
 the length of magnetic loop  is set as $l_{\rm c}\approx R$, which is more reasonable for a geometrically thick, two-temperature hot accretion flows \citep[e.g.][]{1994ApJ...428L..13N, 1995ApJ...444..231N, 1995ApJ...452..710N, 2014ARA&A..52..529Y}  than a constant value of $10R_{\rm S}$ for all distances adopted in previous studies. 
Third, the precision of the energy conservation is substantially improved by implementing the iterative procedures at each localized disc region rather than taking the whole disc as one region in previous studies.
 Fourth,  reflection (for an albedo of 0.2) is taken into account in the iterative calculation. With these refinements, the model is now more self-consistent and strictly energy-conservative (both locally and globally), which can be applied to explain the observational data over a wide energy band from the optical/UV to X-rays.

We point out that  the X-ray emission is dominantly contributed by the innermost corona since it is powered by  the released gravitational energy ($\propto 1/R$), even though the corona extends  to  large distances.  Detailed study of hot corona \citep{2017ApJ...847...96L} shows that the bulk X-ray emissions concentrate in a region within $10R_{\rm S}-20R_{\rm S}$. The disc--corona model is not contradictory with the observed `compact' size measured by X-ray radiation \citep[e.g.][]{2013ApJ...769L...7R, 2015MNRAS.451.4375F, 2016AN....337..356C} .

\subsection{Theoretical spectra}
\label{sec:effects}
In our model, the viscosity parameter $\alpha$ is fixed to the value of $0.3$, as have been done in previous investigations \citep{2003ApJ...587..571L, 2016ApJ...833...35L}. The structure and emergent spectrum of the disc--corona depend on the black hole mass $m$ (in units of solar mass $M_{\sun}$), the mass accretion rate $\dot{m}$ (in units of the Eddington rate $\dot{M}_{\rm Edd}=1.4\times10^{18} \frac{M}{M_{\sun}} {\rm g~s^{-1}}$ for an efficiency $\eta=0.1$) and the magnetic equipartition coefficient $\beta_0$. 
The effects of the parameter $m$, $\dot m$, and $\beta_0$ on the disc--corona structure and further the broad-band spectral features are investigated in details in the following subsections. The main spectral features and their dependence on $m$, $\dot m$, and $\beta_0$ are summarized in the last two paragraphs of this subsection.

\begin{figure*} 
\centering
\includegraphics[width=80mm,height=55mm,angle=0.0]{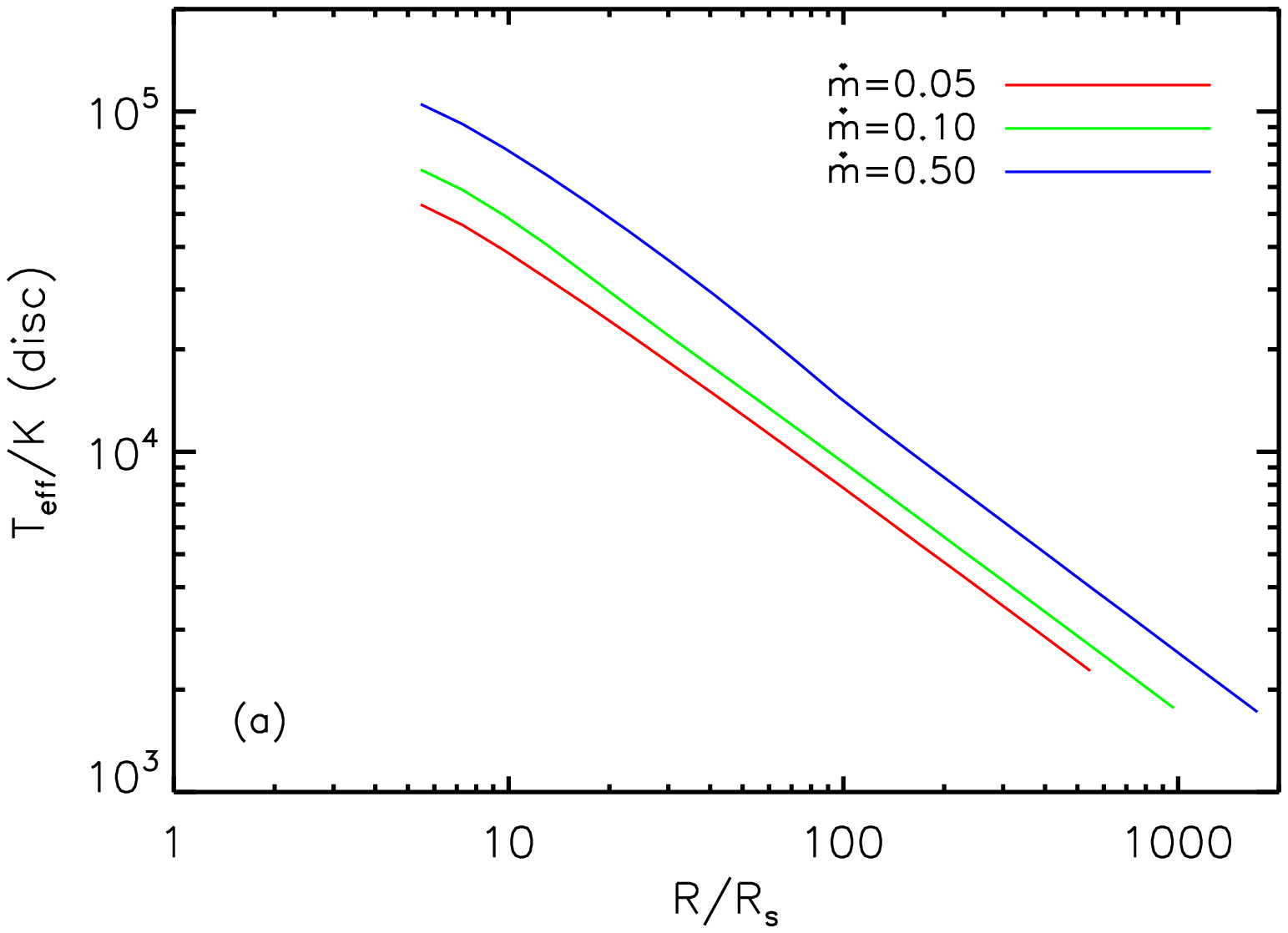}
\includegraphics[width=80mm,height=55mm,angle=0.0]{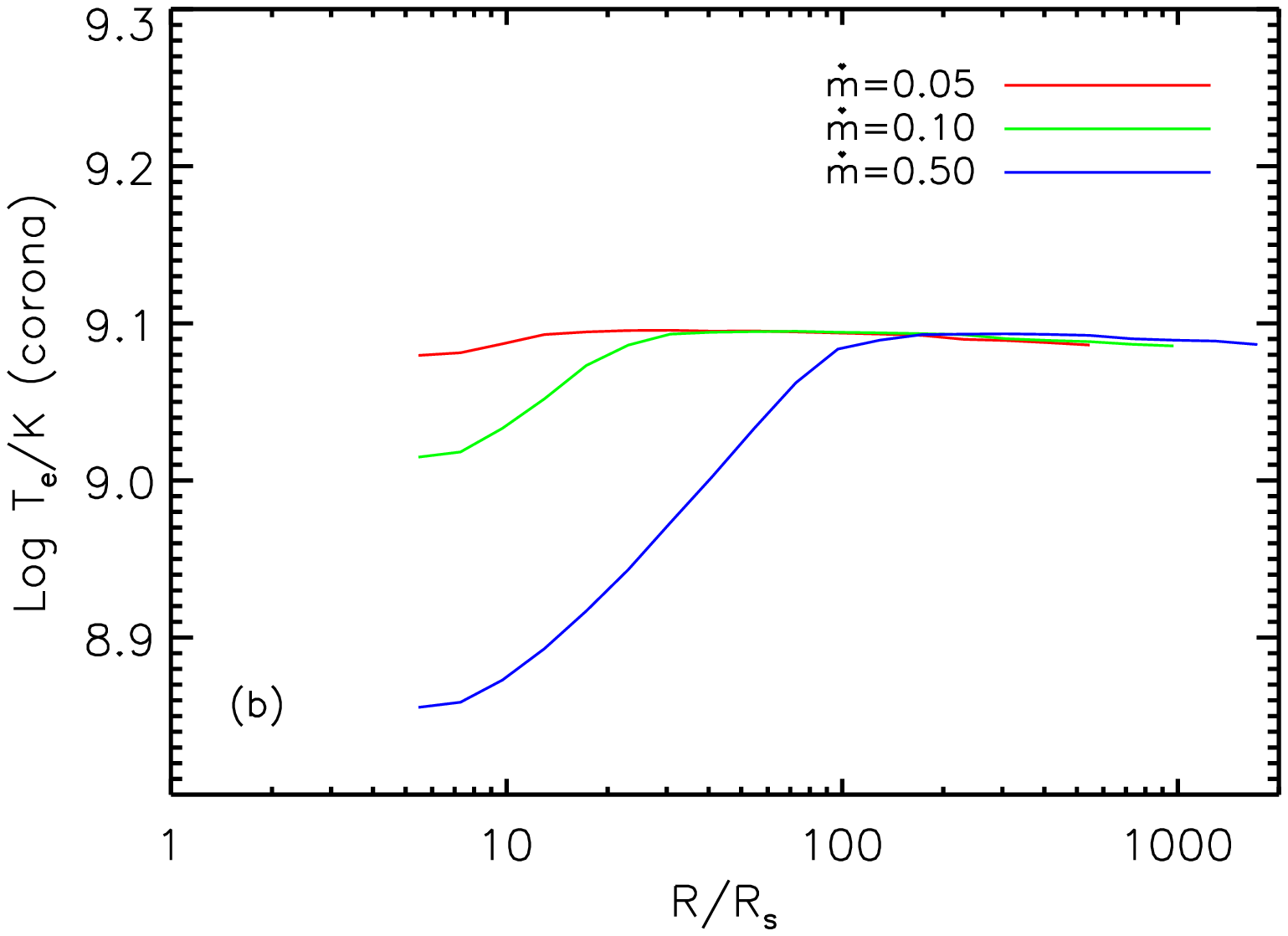}
\includegraphics[width=80mm,height=55mm,angle=0.0]{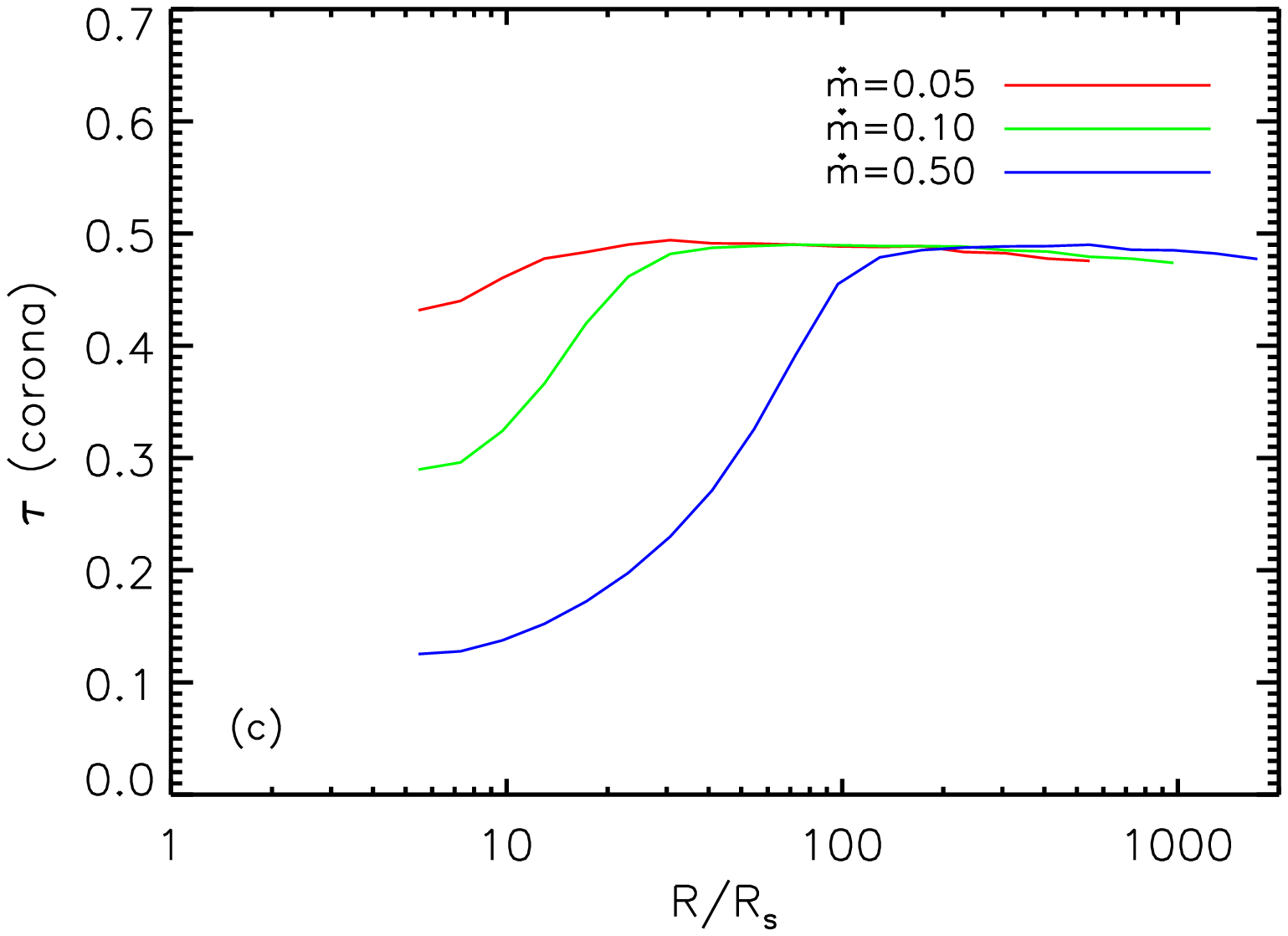}
\includegraphics[width=80mm,height=55mm,angle=0.0]{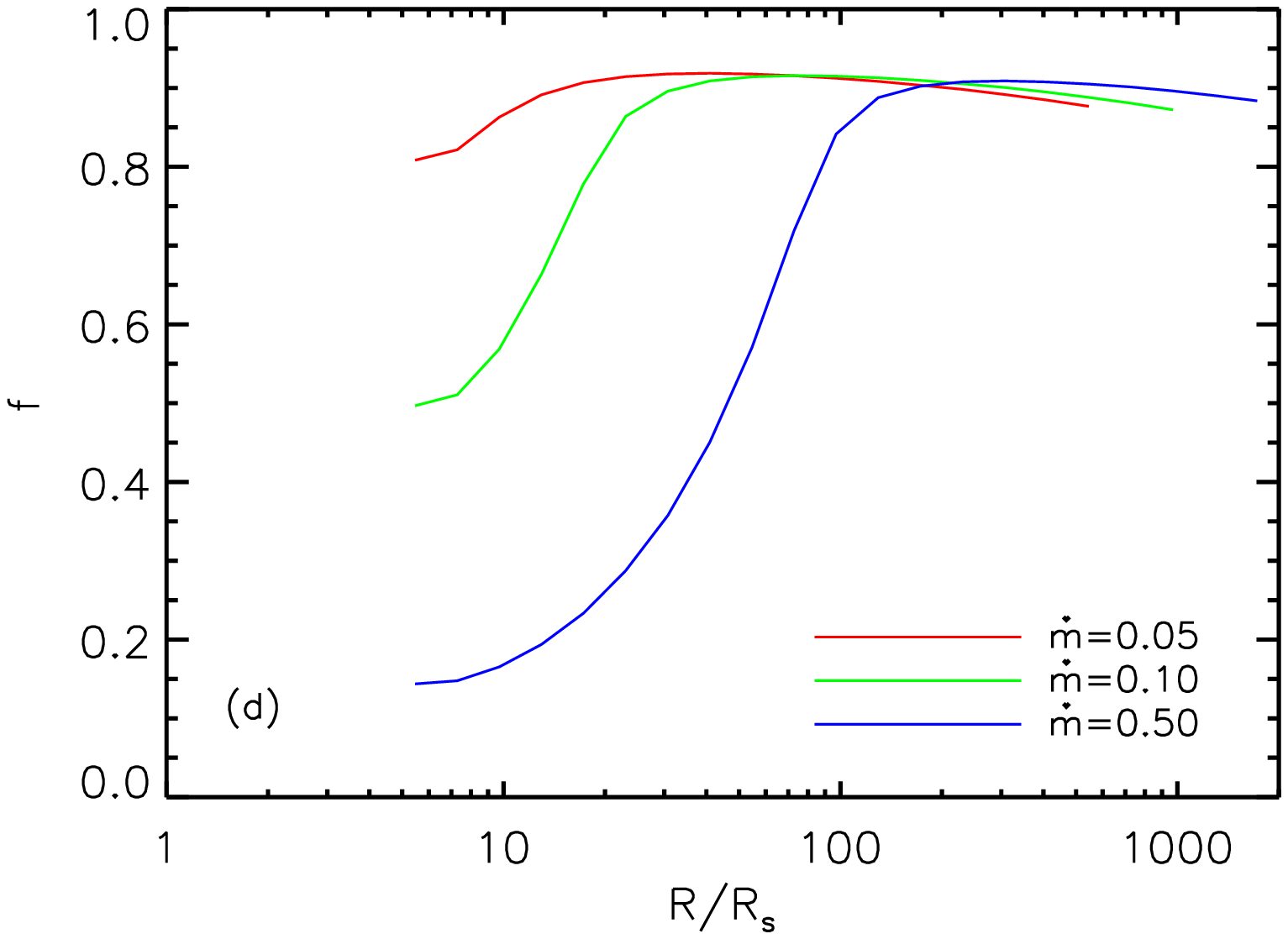}
\caption{Effect of mass accretion rate on the radial distribution of the effective temperature in the thin disc (panel a), the electron temperature in the corona (panel b), the modified optical depth in the corona (panel c), and  the fraction of energy released in the corona  (panel d). The black hole mass and magnetic parameter are fixed to $m=10^8$ and $\beta_0=200$, while the accretion rate changes from 0.05, 0.1, to 0.5. 
The coronal temperature ($T_{\rm e}$), optical depth ($\tau$), and energy fraction ($f$) decrease  dramatically when the disc changes from gas pressure-dominant at large distances to  radiation pressure-dominant  at small distances.}
\label{fig:param_mdot}
\end{figure*}

\begin{figure*} 
\centering
\includegraphics[width=80mm,height=55mm,angle=0.0]{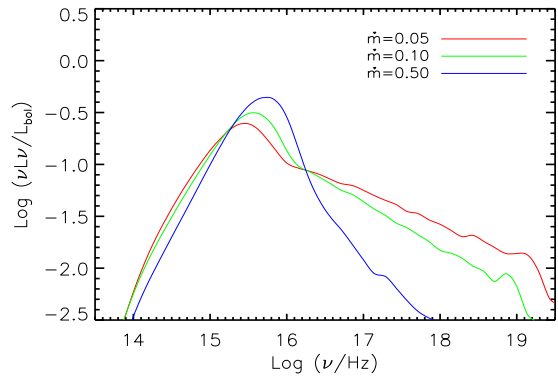}
\caption{Effect of mass accretion rate on the emergent spectra from the disc--corona system. Parameters are the same as Fig.\ref{fig:param_mdot}. The luminosity has been scaled to the bolometric luminosity. 
}
\label{fig:sed_mdot}
\end{figure*}

\subsubsection{The effect of mass accretion rate}
\label{sec:effectmdot}
It has been speculated that the spectrum of an object with given black hole mass is solely determined by the mass accretion rate. The variation of the spectrum is then caused by a change in the accretion rate.  With magnetic field involved in transporting energy from the disc to the corona, an additional parameter, namely the equipartition coefficient $\beta_0$, also plays a role in determining the relative strength of radiations from the disc and the corona. To study the effect of accretion rate, we take values of $m=10^8$ and $\beta_0=200$,  and show the structure of accretion flows and radiation spectra in Fig. \ref{fig:param_mdot} and Fig. \ref{fig:sed_mdot}.

In the panel (a)--(d) of  Fig. \ref{fig:param_mdot}, the radial distributions of effective temperature in the disc, the coronal electron temperature, the coronal modified optical depth, and fraction of the energy released in the corona are plotted for the mass accretion rate $\dot{m}=0.05$ (red), $0.1$ (green), and $0.5$ (blue), respectively. An obvious feature is that the coronal temperature, optical depth, and energy fraction turn over at some distance, with a larger extent at higher accretion rates.  This is because the radiation pressure dominates over the gas pressure in the inner disc region. With the increase of the accretion rate, the radiation pressure-dominant region becomes large, leading to the turn-over at large distance.  As the radiation pressure in the disc solution does not depend on the accretion rates, the strength of magnetic field does not vary with $\dot m$ under the equipartition assumption. Therefore, the energy fraction ($f$) transferred to the corona decreases with increase of accretion energy measured by $\dot m$. Consequently, the disc radiation increases, while the coronal density (equivalent to optical depth)  and temperature are lower due to less heating  (small $f$) and more cooling (caused by strong disc radiation) at higher accretion rates. As shown in panel (b) and (c) of Fig. \ref{fig:param_mdot}, the coronal electron temperature decreases from ${\sim1.3\times10^9}$ to $\sim7\times10^8$ K, and the optical depth decreases from ${\sim0.45}$ to $\sim0.1$ when the accretion rate increases from $0.05$ to $0.5$. We note that the radial distribution of the effective temperature is not similar to that of corona.  This is because roughly half of the corona radiation goes back to illuminate the disc, no matter how much accretion energy is transferred to the corona.  The inclusion of illumination could also lead to a quantitative deviation from the above analyses but could not change the variation trend with $\dot m$. 

The spectrum is determined by the structure of the two-phase accretion flow.  With the dependence of the disc--corona structure on the accretion rate, we expect that the ratio of X-ray emission to disc decreases with increase of $ \dot m$ as a result of decrease in energy fraction $f$; The optical/UV spectrum shifts to higher frequencies due to more accretion energy released in the disc; The decrease of coronal temperature and optical depth  leads to a large decrease of Compton $y$-parameter, thereby  significantly softens the X-ray spectrum. The detailed spectral index from IC scattering in optically thin corona is $\alpha\approx{\ln{1\over\tau_{\rm es}}/{\ln{A}}}$,
where the amplification factor $A$ is determined by electron temperature, $A=1+4\theta_{\rm e}+16 {\theta_{\rm e}}^2$ with $\theta_{\rm e}= {kT_{\rm e}\over {m_{\rm e}c^2}}$. Such predictions are confirmed by our computational results, as illustrated in Fig. \ref{fig:sed_mdot}. That is, from $\dot m=0.05$ to $0.5$, the X-ray spectrum becomes weaker and softer, and the disc component becomes stronger and shifts to higher frequency. Such a feature is in agreement with observations in AGN \citep[e.g.][]{2006ApJ...646L..29S, 2007MNRAS.381.1235V, 2012MNRAS.425..907J}. Nevertheless, the obtained spectra as shown in Fig. \ref{fig:sed_mdot} are still too soft compared to the observational X-ray spectra in some AGNs.   A stronger magnetic field is expected to produce 
a harder X-ray spectrum for a given accretion rate, 
which is discussed in the following subsection.

\begin{figure*} 
\centering
\includegraphics[width=80mm,height=55mm,angle=0.0]{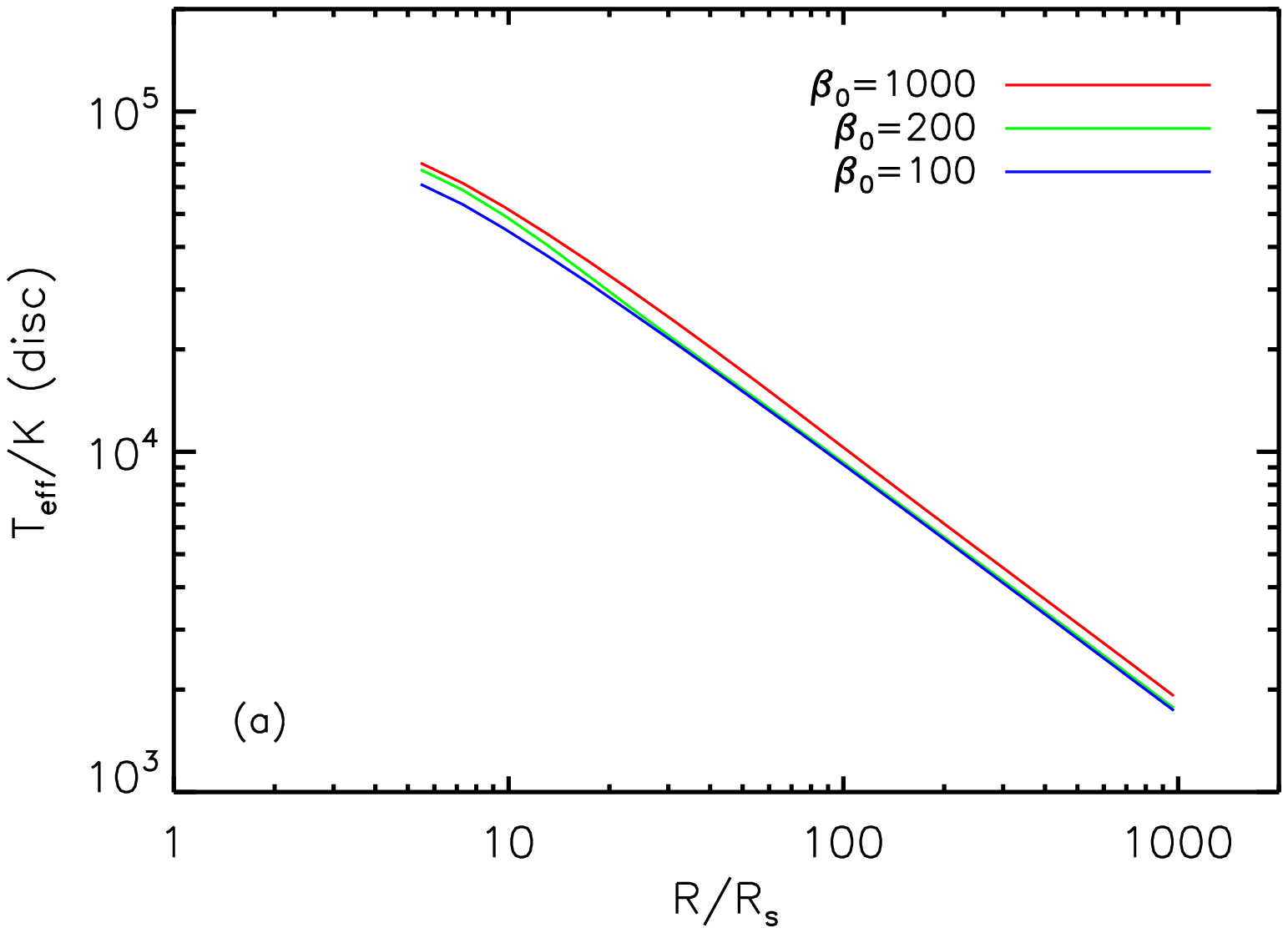}
\includegraphics[width=80mm,height=55mm,angle=0.0]{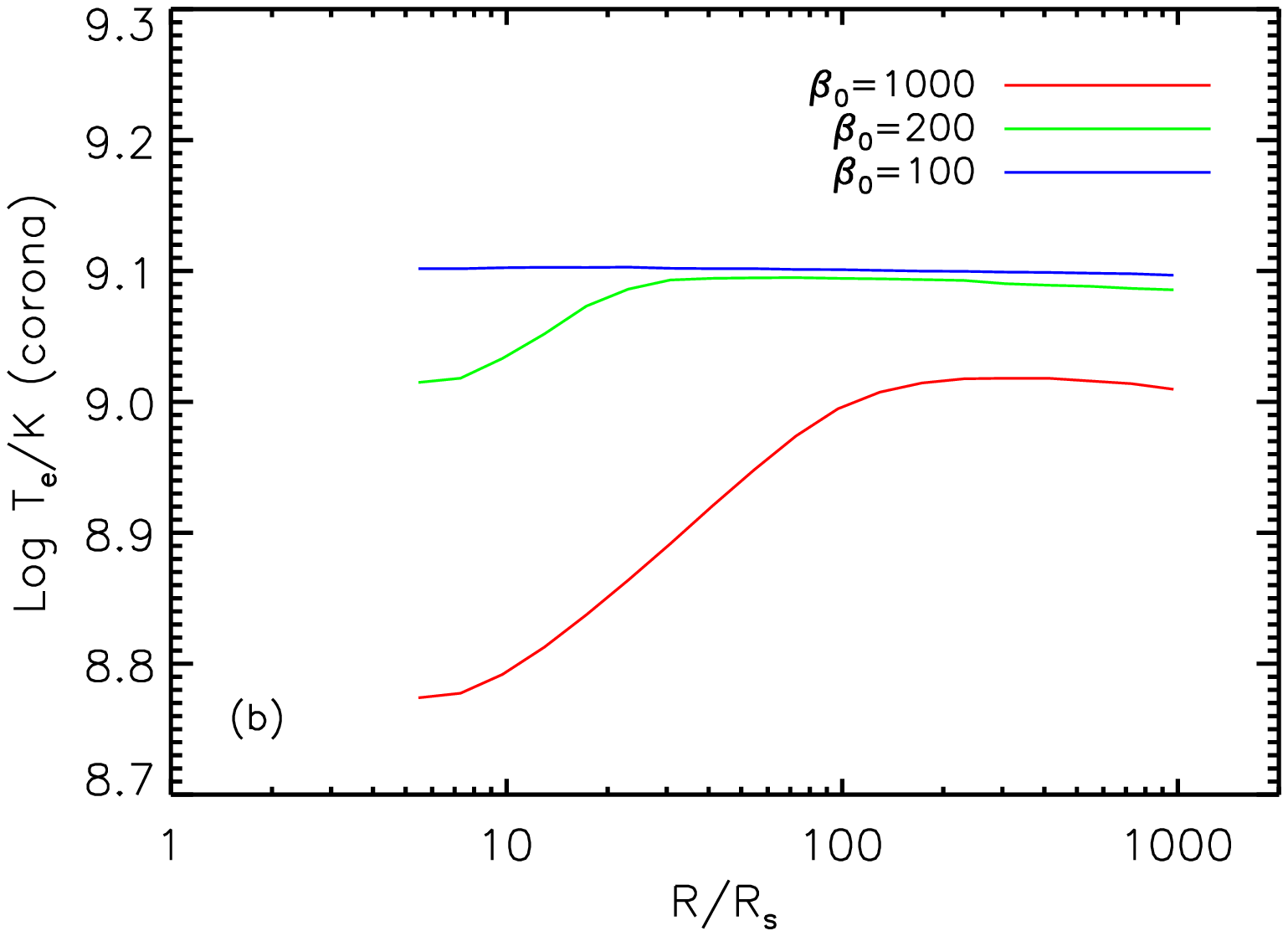}
\includegraphics[width=80mm,height=55mm,angle=0.0]{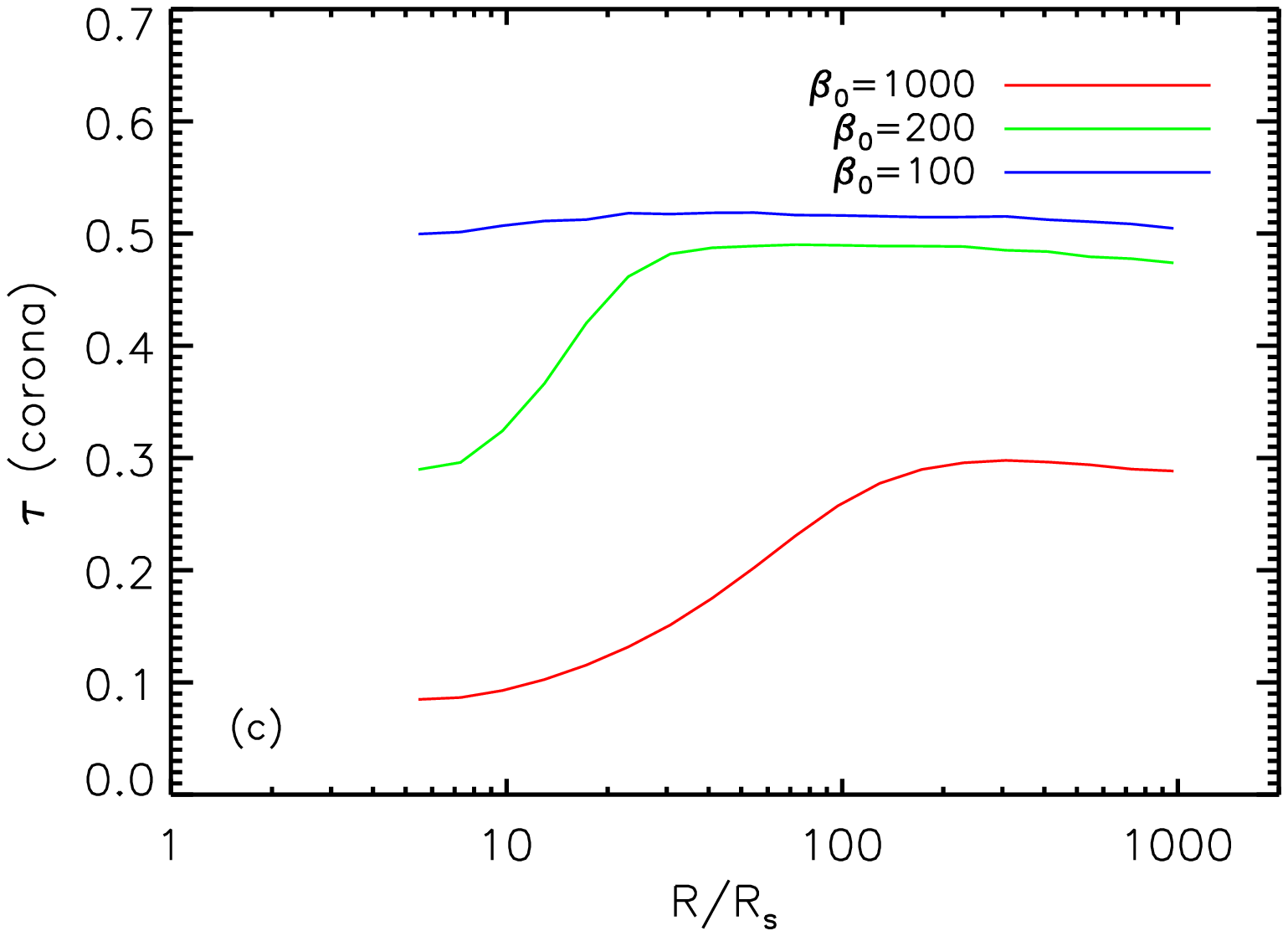}
\includegraphics[width=80mm,height=55mm,angle=0.0]{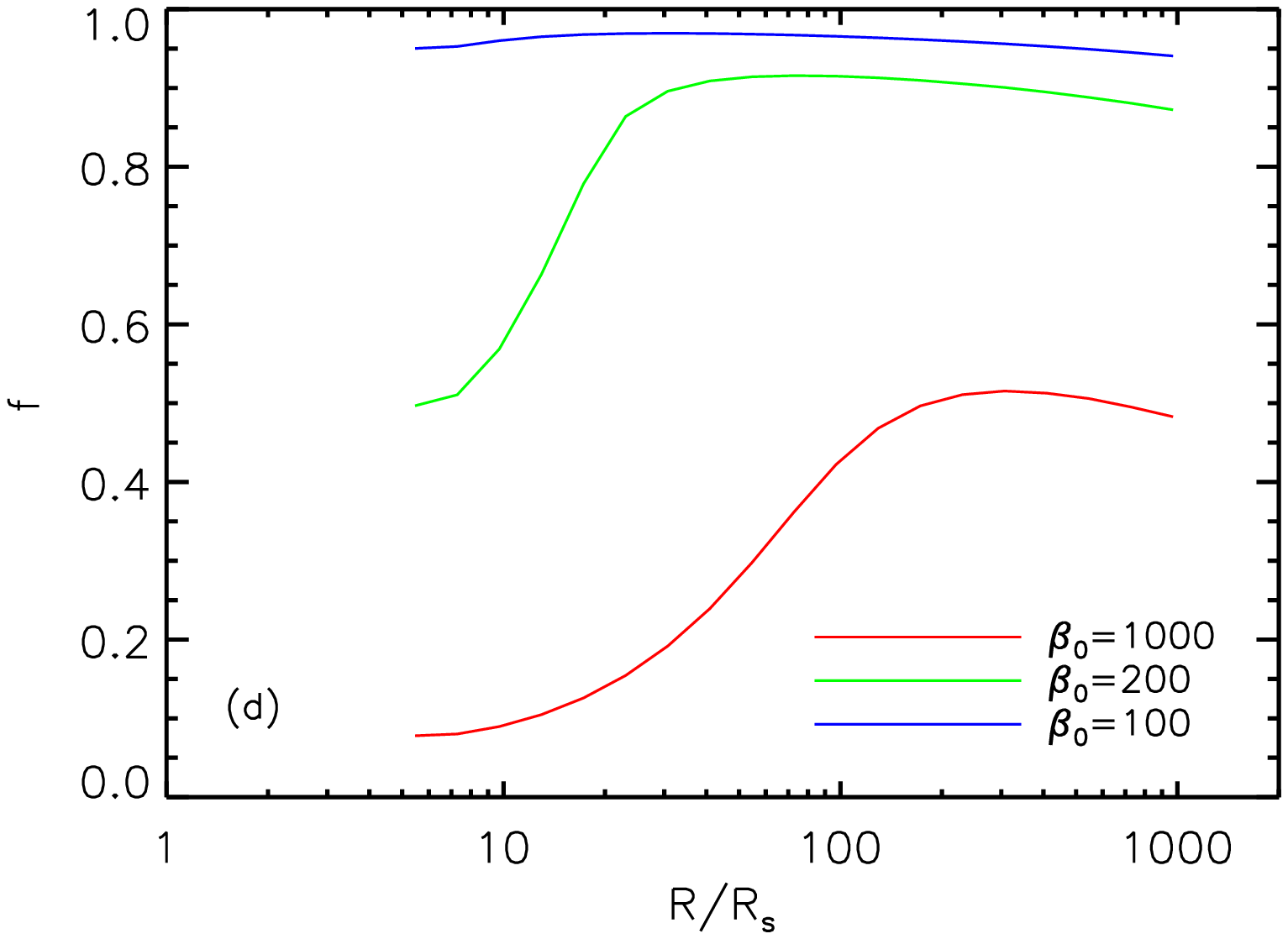}
\caption{Effect of magnetic field on the radial distribution of the effective temperature in the thin disc (panel a), the electron temperature in the corona (panel b), the modified optical depth in the corona (panel c), and  the fraction of energy released in the corona  (panel d). The black hole mass and magnetic parameter are fixed to $m=10^8$ and $\dot m=0.1$, while the magnetic parameter changes from 100, 200, to 1000. 
The coronal temperature ($T_{\rm e}$), optical depth ($\tau$), and energy fraction ($f$) decrease  dramatically when the disc changes from gas pressure-dominant at large distances to  radiation pressure-dominant  at small distances.}
\label{fig:param_beta}
\end{figure*}

\begin{figure*} 
\centering
\includegraphics[width=80mm,height=55mm,angle=0.0]{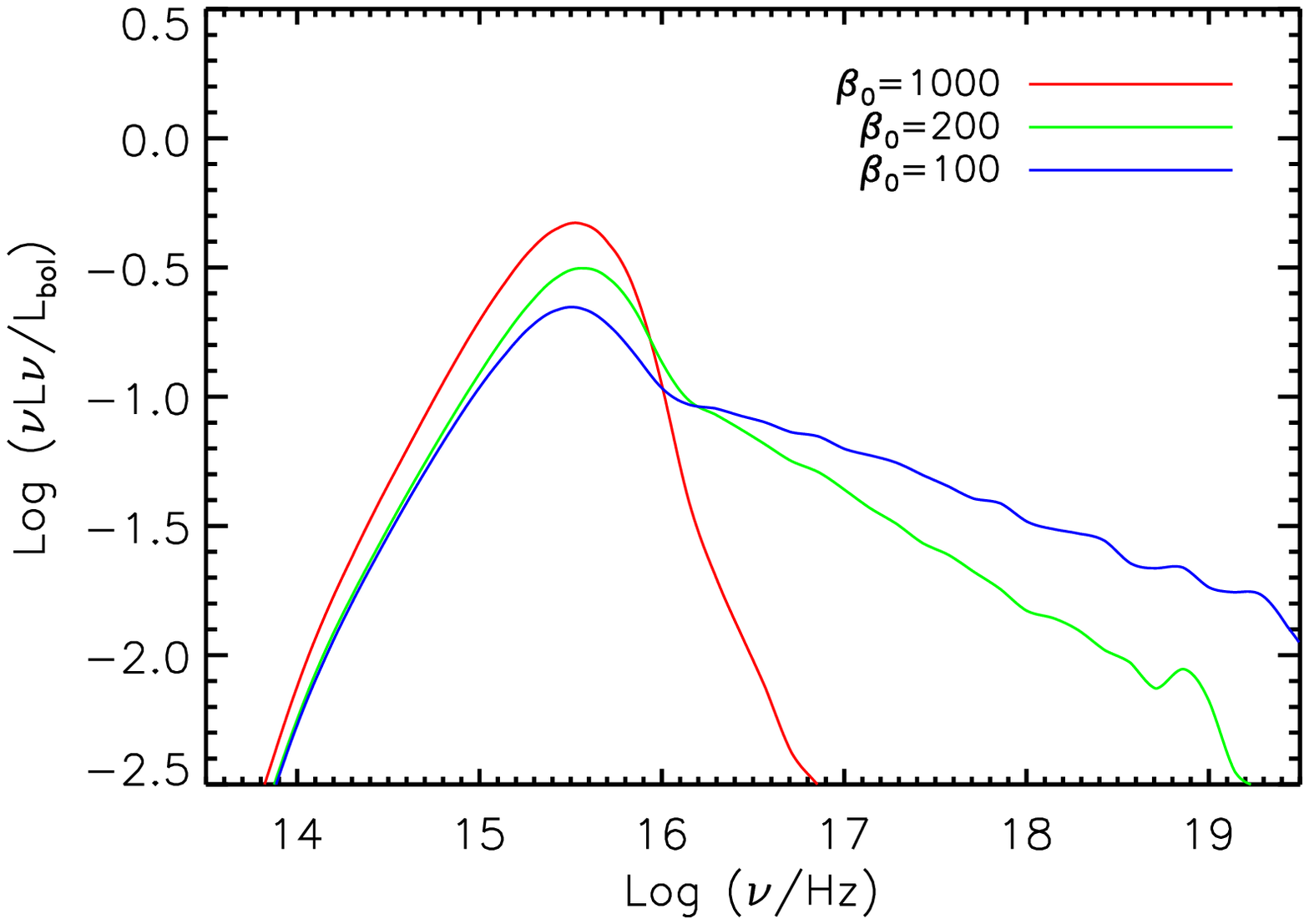}
\caption{ Effect of magnetic field on the emergent spectra from the disc--corona system. Parameters are the same as Fig.\ref{fig:param_beta}.  The luminosity has been scaled to the bolometric luminosity. }
\label{fig:sed_beta}
\end{figure*}

\subsubsection{The effect of the strength of magnetic field}
\label{sec:effectbeta}

The magnetic equipartition coefficient, $\beta_0$, is defined as the ratio of the sum of the gas and radiation pressure to the magnetic pressure. A large value of  $\beta_0$ implies a weak magnetic field. For different $\beta_0$ values, we calculate the disc--corona structure and then the emergent spectrum. The radial distributions of effective temperature in the disc, the coronal electron temperature, the coronal modified optical depth, and the fraction of the energy released in the corona are plotted in Fig. \ref{fig:param_beta} for fixed $m=10^8$, $\dot{m}=0.1$, and different magnetic equipartition coefficient $\beta_0=100$ (blue), $200$ (green), and $1000$ (red).  The emergent spectra are shown in Fig. \ref{fig:sed_beta}. It can be seen that the increase of the magnetic field (equivalent to a decrease in $\beta_0$) results in an increase in the coronal temperature, optical depth (density),  and hence strong X-ray radiation,  whilst the disc emission decreases.  This is easy to understand since a stronger magnetic field means more energy can be transported into the corona, which evaporates more gas to a higher temperature and produces stronger Compton radiations.  In this manner, the magnetic field can significantly affect the coronal temperature and density, and thereby changes the spectrum significantly,  in particular,  the hard X-ray spectral shape.  In contrast, the effective temperature of the disc depends only weakly on $\beta_0$ because the energy released in the corona partially comes back to heat the disc through irradiation. 

We find that for $\beta_0=100$ and $\dot{m}=0.1$, nearly all the accretion energy is transferred to the corona by magnetic field  ($f\approx1$), and the disc is heated by the coronal irradiation rather than viscosity, which is referred to a passive disc.  Further increase of magnetic field does not change the results. We note that critical value of $\beta_0$ for becoming a passive disc ($f\approx1$)  depends on the mass accretion rates. For instance, the passive disc solution ($f\approx1$) is satisfied when $\beta_0$ decreases to $\sim50$  at an accretion rate of $\dot{m}=0.5$, compared to $\beta_0=100$ at $\dot m=0.1$; However, in the case of $\dot{m}=0.01$, this can be achieved for $\beta_0=200$. 

The magnetic field also affects the radial structure of disc and corona, as shown in the radial distribution of coronal temperature and optical depth in Fig. \ref{fig:param_beta}.  In the case of strong magnetic field, the disc can be gas pressure-dominant at almost all radii since a large amount of energy is transferred to the corona (say, $\beta_0=100$); while in the case of weak magnetic field ($\beta_0=1000$), only ${\sim 16}$ per cent of the accretion energy is released in the corona, and the radiation pressure-dominant region in the disc extends to a distance of $R\sim 100R_{\rm S}$. The size of the radiation pressure-dominant region in the disc leads to a corresponding change of  the coronal temperature and optical depth  along distance, as discussed in above subsection.

\begin{figure*} 
\centering
\includegraphics[width=80mm,height=55mm,angle=0.0]{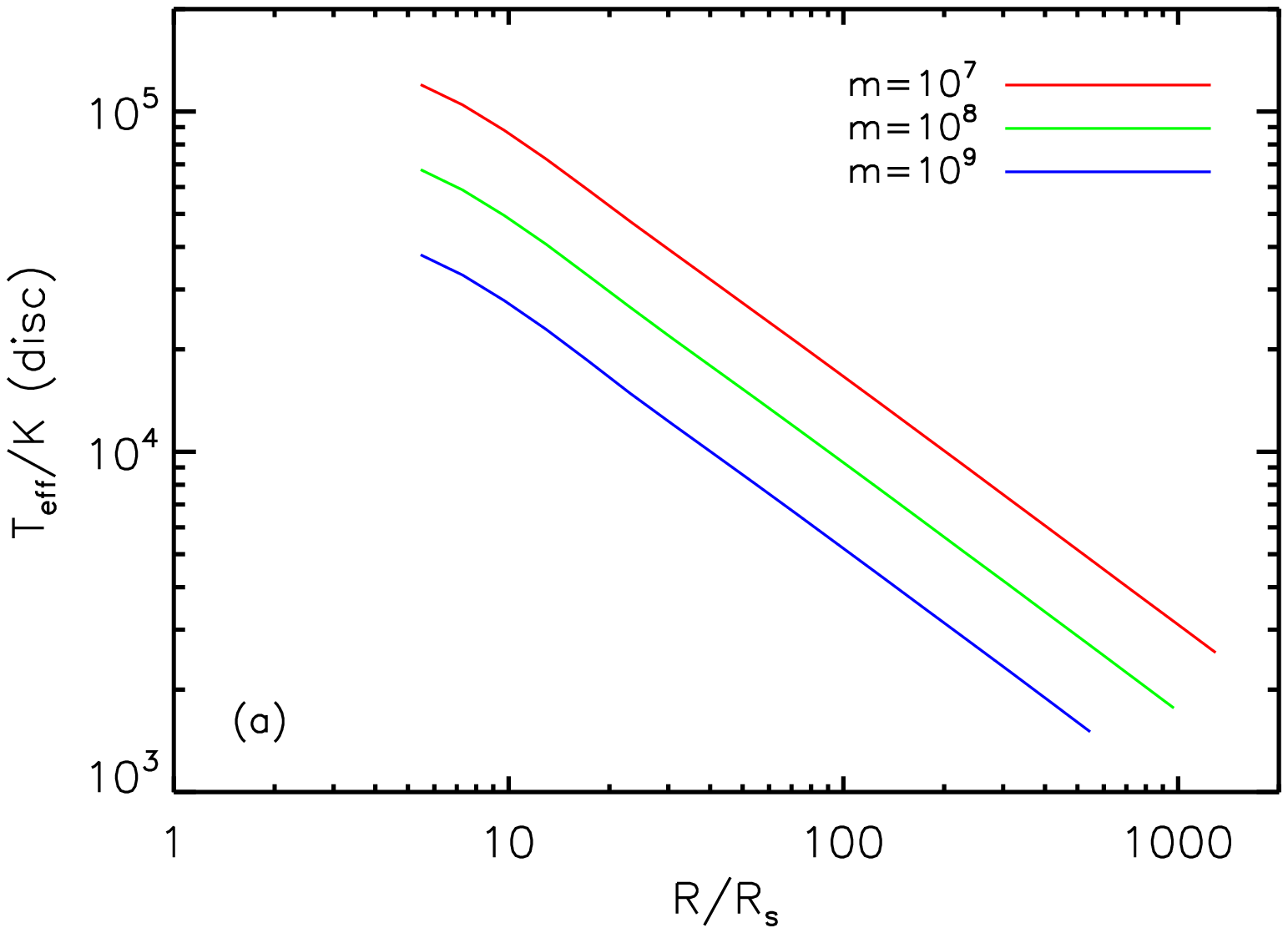}
\includegraphics[width=80mm,height=55mm,angle=0.0]{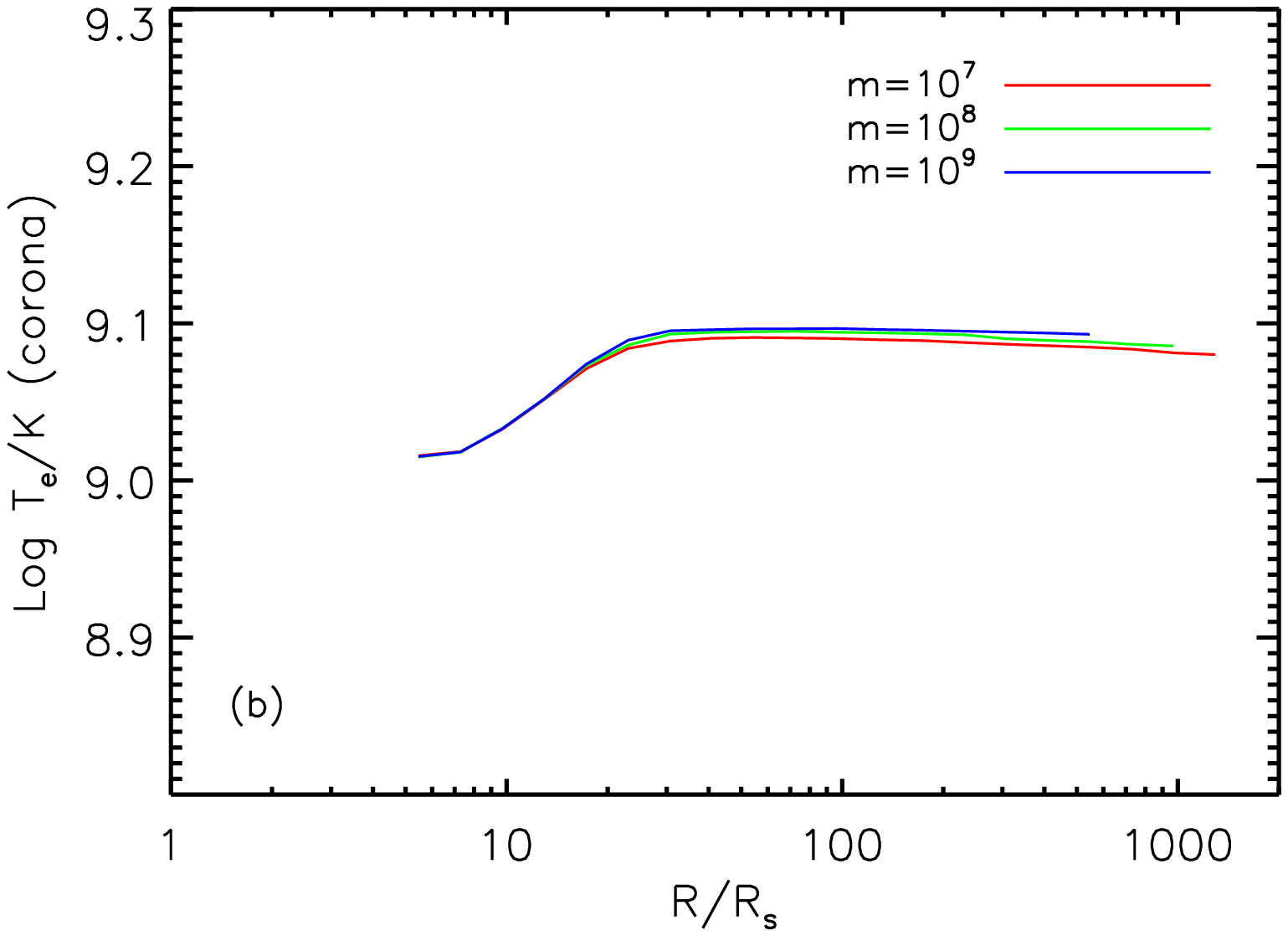}
\includegraphics[width=80mm,height=55mm,angle=0.0]{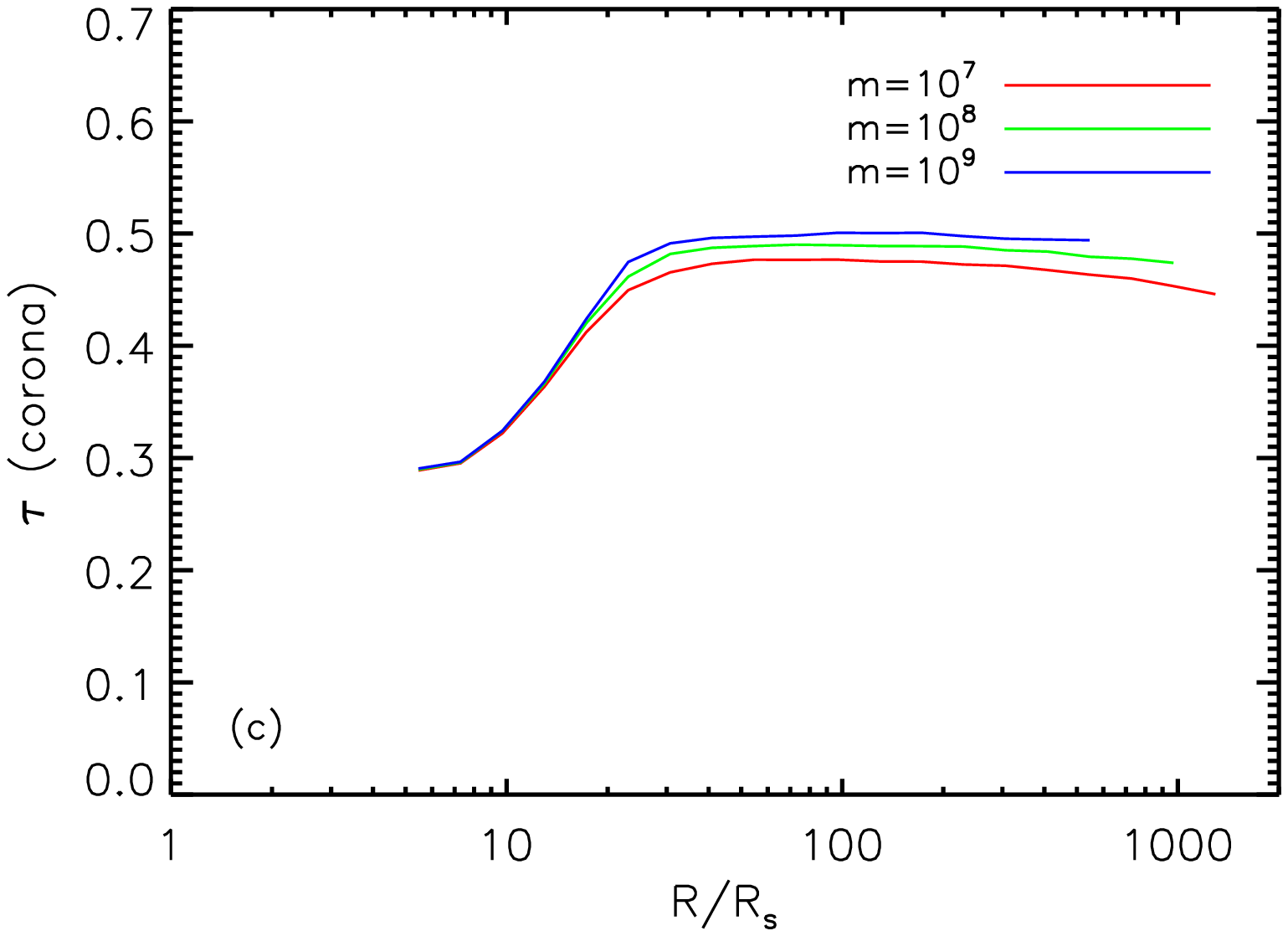}
\includegraphics[width=80mm,height=55mm,angle=0.0]{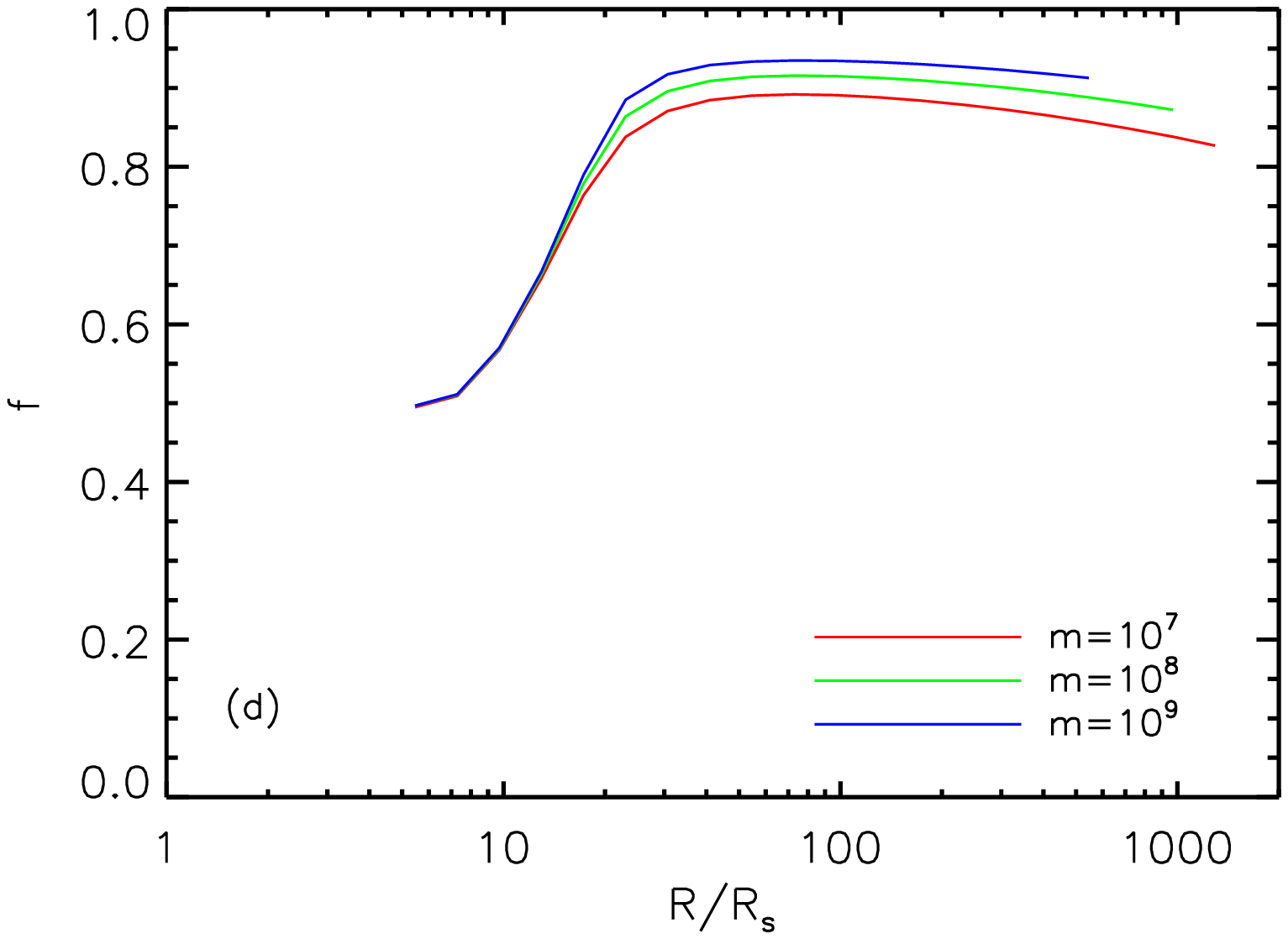}
\caption{Effect of black hole mass on the radial distribution of the effective temperature in the thin disc (panel a), the electron temperature in the corona (panel b), the modified optical depth in the corona (panel c), and  the fraction of energy released in the corona  (panel d). The accretion rate and magnetic parameter are fixed to  $\dot m=0.1$ and $\beta_0=200$, while the black hole mass changes from $10^7$, $10^8$, to $10^9$ solar mass. 
The coronal temperature ($T_{\rm e}$), optical depth ($\tau$), and energy fraction ($f$) decrease  dramatically when the disc changes from gas pressure-dominant at large distances to  radiation pressure-dominant  at small distances.}
\label{fig:param_mass}
\end{figure*}

\begin{figure*} 
\centering
\includegraphics[width=80mm,height=55mm,angle=0.0]{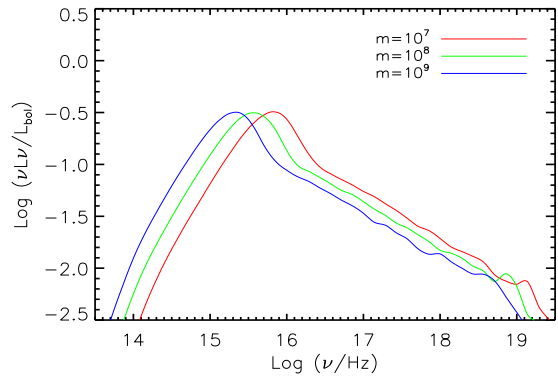}
\caption{ Effect of black hole mass on the emergent spectra from the disc--corona system. Parameters are the same as Fig.\ref{fig:param_mass}. The luminosity has been scaled to the bolometric luminosity. }
\label{fig:sed_mass}
\end{figure*}

\subsubsection{The effect of black hole mass}
\label{sec:effectmass}

The black hole mass in AGN spans typically from $10^6$ to $10^9$ solar mass. This can lead to significant differences in the effective temperature of a standard disc  ($T_{\rm eff}\propto m^{-1/4}$) for individual AGN.  We study the influence of black hole mass on both the disc and corona structure and the emergent spectrum.
In the panels (a)--(d) of Fig. \ref{fig:param_mass}, we plot the effective temperature in the disc, the electron temperature, the modified optical depth and the energy fraction in the corona as a function of radius for the black hole mass $m=10^7$ (red), $10^8$ (green), and $10^9$ (blue), with given $\dot{m}=0.1$ and $\beta_0=200$. Similar to the standard Shakura--Sunyaev disc, the effective temperature of the disc is lower for a higher black hole mass, and the radiation shifts to lower frequencies as shown in Fig. \ref{fig:sed_mass}. On the other hand, the effects of the black hole mass on the distribution of coronal electron temperature $T_{\rm e}$ and modified optical depth $\tau$ can nearly be neglected. This is a common nature of optically thin, hot accretion flows, no matter it is an ADAF, a friction-heated corona, or magnetic-reconnection-heated corona.  Consequently, the hard X-ray spectral index is independent of the black hole mass.  Combined with the effect of $m$ on the disc radiation, the SED from disc and corona shifts to lower frequency for higher black hole mass, as shown in Fig. \ref{fig:sed_mass}. 

To summarize, the theoretical spectrum from the disc and magnetic-reconnection-heated corona consists of two primary portions, that is the optical-to-EUV radiation is dominated by the multicolor blackbody emission from optically thick disc, while the radiation from EUV to hard X-ray band is dominated by the Comptonization process of the disc photons, forming a power-law spectrum in the hard X-ray band below $\sim100$ keV or so. The emergent spectrum from the disc--corona varies with three physical parameters inherent in the model, that is the black hole mass $m$, the mass accretion rate $\dot{m}$, and the magnetic equipartition coefficient $\beta_0$. The X-ray spectrum is steeper at higher accretion rate, implying a stronger disc (relative to the corona) when the source is brighter.  Similarly, the X-ray emission increases with the increase of magnetic field, meanwhile the disc emission decreases. Compared to the effect of accretion rate, the spectral features in the X-ray band are more sensitive to the variations in $\beta_0$, since it directly determines the energy released in the corona. The black hole mass only affects the disc component. The higher the black hole mass is, the lower frequency of disc radiation peaks at.

Such a model provides a physical mechanism to interpret the variety of observational AGN spectra with limited free parameters. In particular, the magnetic buoyant instability and magnetic-reconnection scenario solves the energy shortage of the corona, making it possible to produce sufficiently strong X-ray emission as observed in AGN. The energy fraction liberated in the corona, $f$, is self-consistently determined by the magnetic behavior and the coupling of the disc and the corona, which was usually taken as a free parameter without a physical mechanism in previous studies. The predicted spectral features are in qualitative agreement with observational characteristics in AGN. Modelling of observed SED for a sample of AGN is performed in the next section.   

\section{Modelling the observed broad-band SED of AGN}
\label{sec:modelsed}

On the basis of above theoretical investigations, we are ready to model the observational spectra of AGN from just the two fundamental parameters -- the black hole mass and accretion rate -- plus the magnetic equipartition coefficient. In the following spectral fitting, $m$, $\dot m$, and $\beta_0$ are set to be as free parameters, which can be constrained from the shapes and luminosities of the observed SEDs of individual AGN.

\subsection{Sample and data}
\label{sec:sample}

In a recent study, we obtained the broad-band SEDs of a sample of $23$ local AGNs \citep{2019MNRAS.487.3884C}.  The sample is adopted for spectral modelling in this investigations given the following advantages. First, the optical, UV, and X-ray data are obtained by simultaneous observations with the UV/Optical telescope \citep[UVOT,][]{2005SSRv..120...95R} and X-ray telescope \citep[XRT,][]{2005SSRv..120..165B} onboard the Neil Gehrels {\it Swift} Observatory \citep{2004ApJ...611.1005G}, which are essential for SED studies. Moreover, the sample was selected to be largely free from dust reddening using the Balmer decrement as an indicator. Secondly, photometry of the nuclei was measured in an elaborate way by eliminating the host galaxy starlight. Thirdly, their optical spectral parameters were measured accurately in a homogenous way by taking advantage of the Sloan Digital Sky Survey \citep[SDSS,][]{2000AJ....120.1579Y} spectroscopic observations of all the sample objects \citep[see][for details]{2019MNRAS.487.3884C}. Therefore, the virial black hole masses ($M_{\rm vir}$) were reliably measured from the single-epoch spectral parameters, and thus the Eddington ratios for the sample objects.

The luminosities of the sample objects in the optical, UV, and X-ray bands are taken from \citet{2019MNRAS.487.3884C}. It should be noted that the optical/UV luminosities are derived from the measured photometric fluxes for disc emission assuming an inclination angle of $30^{\circ}$ of the disc normal with respect to the line of sight \citep[see][Section 4.1]{2019MNRAS.487.3884C}; whereas the X-ray luminosities are derived assuming isotropic emission of the optically thin corona. To facilitate the modelling of the SED, the uncertainties of some of the measurements in \citet{2019MNRAS.487.3884C} are  estimated or refined. First, the errors on the measurement of the luminosities in optical/UV band are quantitively reevaluated. For the error on the measurement of the luminosities in the optical ({\it u, b, v}) band, we include the standard measurement error, the systematic error due to the calibration and an extra error of $20$ per cent to take into account the uncertainty in performing the AGN--host galaxy decomposition, which is estimated via the comparison of our results with those of \citet{2011ApJ...739...57K} for the seven same objects in the two samples. For UV ({\it uvw1, uvm2,} and {\it uvw2}) bands, the standard error from Poisson noise is combined with the flux calibration error and an additional error of $10$ per cent on the uncertainty associated with the possible host galaxy contribution, which is estimated based on the results in \citet{2009MNRAS.399.1553V}. The uncertainty associated with the estimation of the dust extinction effect is not considered as the sample is selected in a way that there is little or no intrinsic dust reddening as indicated by the Balmer decrement of the broad emission lines.

It should be noted that the soft X-ray data ($0.3$--$2$ keV) is not taken into account in the spectral fitting. This is because this energy band is often contributed by the so-called soft X-ray excess, whose origin is yet unknown \citep[e.g.][]{1998MNRAS.301..179M, 2012MNRAS.420.1848D} and whose contribution to the total accretion energy budget is negligible \citep[see][for details]{2019MNRAS.487.3884C}. Therefore, only the X-ray spectra in the $2$--$10$ keV band and the optical/UV data are utilized. 
In some of the objects the 2--10 keV spectra are of low quality due to 
the relatively low effective area of XRT above 2\,keV.

Since we consider our disc--corona model to be applicable in the range of $\dot{m}$, that is accretion rate in units of the Eddington rate, from $0.01$ to $1$ (see below), three of the sample objects with significantly lower Eddington ratios ($\lambda_{\rm Edd}<<0.01$) are excluded. Thus the final sample used in the following data modelling comprises only $20$ objects, whose basic data are listed in Table \ref{tab:params} and their broad-band SEDs are shown in Fig. \ref{fig:broadseds}.

\subsection{Spectral fitting procedure}

The fitting procedure is implemented by comparing the observed broad-band spectra with theoretical disc--corona SEDs. For this purpose, we construct a grid of models covering wide ranges in the parameter space of the black hole mass, mass accretion rate and magnetic equipartition coefficient. 
Specifically, a grid of ${496~400}$ models is constructed, 
with the parameter ranges of 
$6.5$--$9.4$ for $\log{(M/{\rm M_{\sun}})}$, 
$0.01$--$1.0$ for $\dot{m}$ and 
$0.001$--$0.034$ for $1/\beta_0$ (see Table \ref{tab:tablemodel}). 
The range of black hole mass is chosen based on the distribution 
of the virial masses of the sample considering 
their typical systematic uncertainty of $0.5$\,dex \citep{2017ApJ...851...21G}.
We note that for each individual AGN, 
the black hole mass is allowed to vary within 
an uncertainty range of $\pm 0.5$\,dex of the virial mass only,
if the fitted mass value (with its uncertainty taken into account)
exceeds significantly the allowed range.
The range of the mass accretion rate of $0.01$--$1.0$ is chosen 
to validate the thin disc assumption, 
beyond which the thin disc could be replaced by an ADAF \citep[e.g.][]{1977ApJ...214..840I, 1982Natur.295...17R, 1995ApJ...444..231N, 1995ApJ...452..710N, 1995ApJ...438L..37A, 2014ARA&A..52..529Y} or a slim disc \citep[e.g.][]{1988ApJ...332..646A, 1999ApJ...522..839W, 2002ApJ...574..315O}. 
Here the critical accretion rate for a slim disc to occur is assumed to be $1$ (rather than $0.3$) since part of the accreted energy is carried away by the magnetic field. The range of the magnetic equipartition coefficient $\beta_0$ is set in such a way that the energy fraction $f$ varies from $0$ to $1$. 

For each set of the model parameters, 
the $\chi^2$ statistic is calculated using the SED data.
The best fit is obtained by searching for the global minimum of 
the resulting $\chi^2$ grid over the 
entire parameter space specified as in Table \ref{tab:tablemodel}.
The confidence range of a given parameter 
at the 90 per cent level for one interesting parameter
is derived as the boundaries around the best-fitting value
beyond which the $\chi^2$ increment $\Delta \chi^2 \ge 2.71$
(with the other parameters freely fitted).
To assess the statistical goodness-of-fit, the null hypothesis probability $p$ 
of the best fit is also derived from the $\chi^2$ distribution.

\begin{table}
\centering
\caption{Ranges and steps of the parameters adopted to construct the grid of models used for SED fitting}
\label{tab:tablemodel}
\begin{threeparttable}
\small
\renewcommand\arraystretch{1.0}
\begin{tabular}{ccccccccccc} 
\hline
\hline
Parameter & Min--Max values & Step & Number of steps\\
\hline
$\log ~(M/{\rm M_{\sun}})$ & $6.5-9.4$ & ${0.02}$ & ${146}$ \\
$\dot{m}$ & $0.01-1.0$ & $0.01$ & $100$ \\
$1/\beta_0$ & $0.001-0.034$ & $0.001$ & $34$ \\
\hline
\end{tabular}\end{threeparttable}
\end{table}

\subsection{Fitting results}
\label{sec:sedresult}

\begin{figure*}
	\vspace{1cm}
	\includegraphics[width=5.7cm]{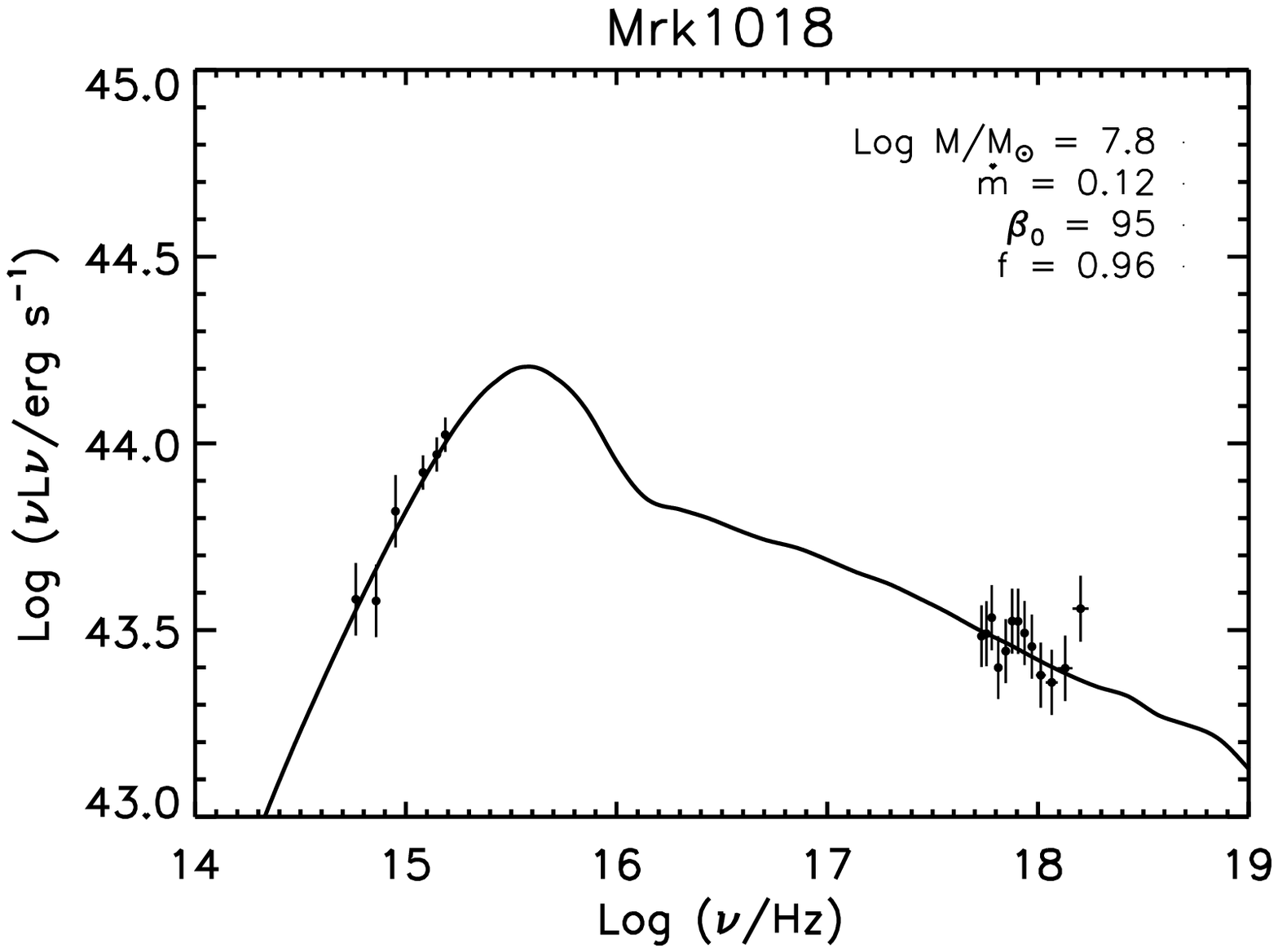}
	\includegraphics[width=5.7cm]{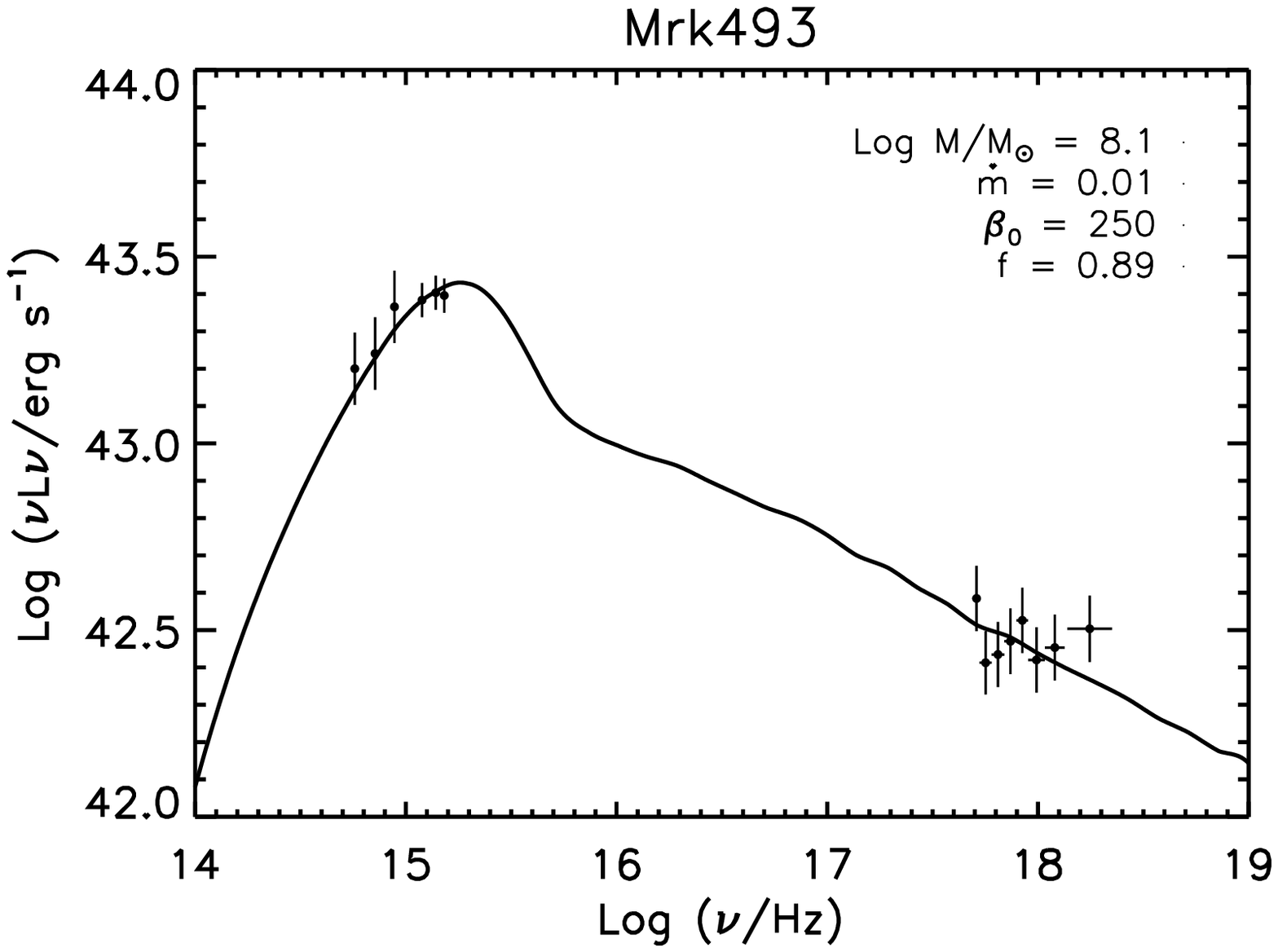}
	\includegraphics[width=5.7cm]{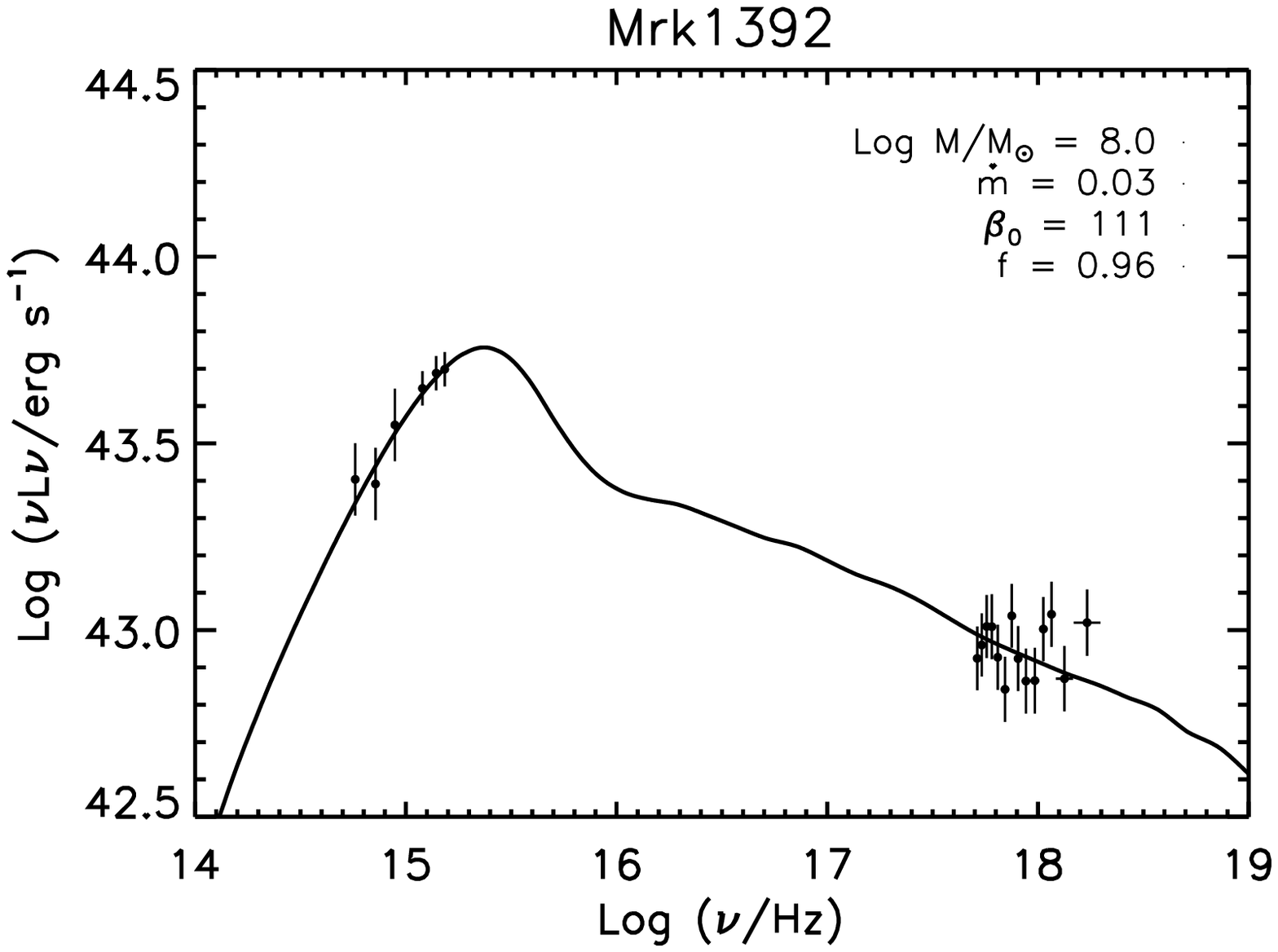}
	\includegraphics[width=5.7cm]{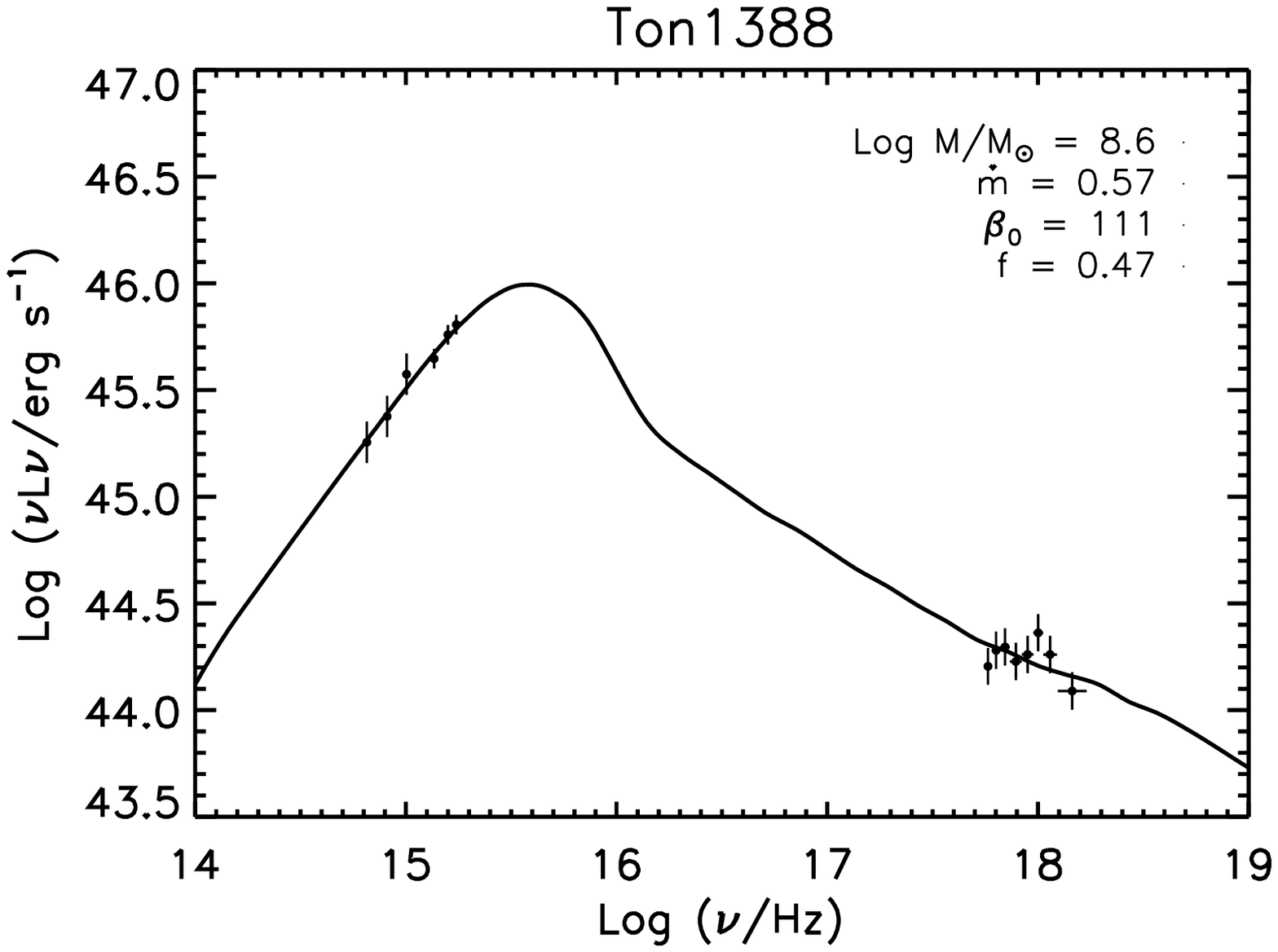}
	\includegraphics[width=5.7cm]{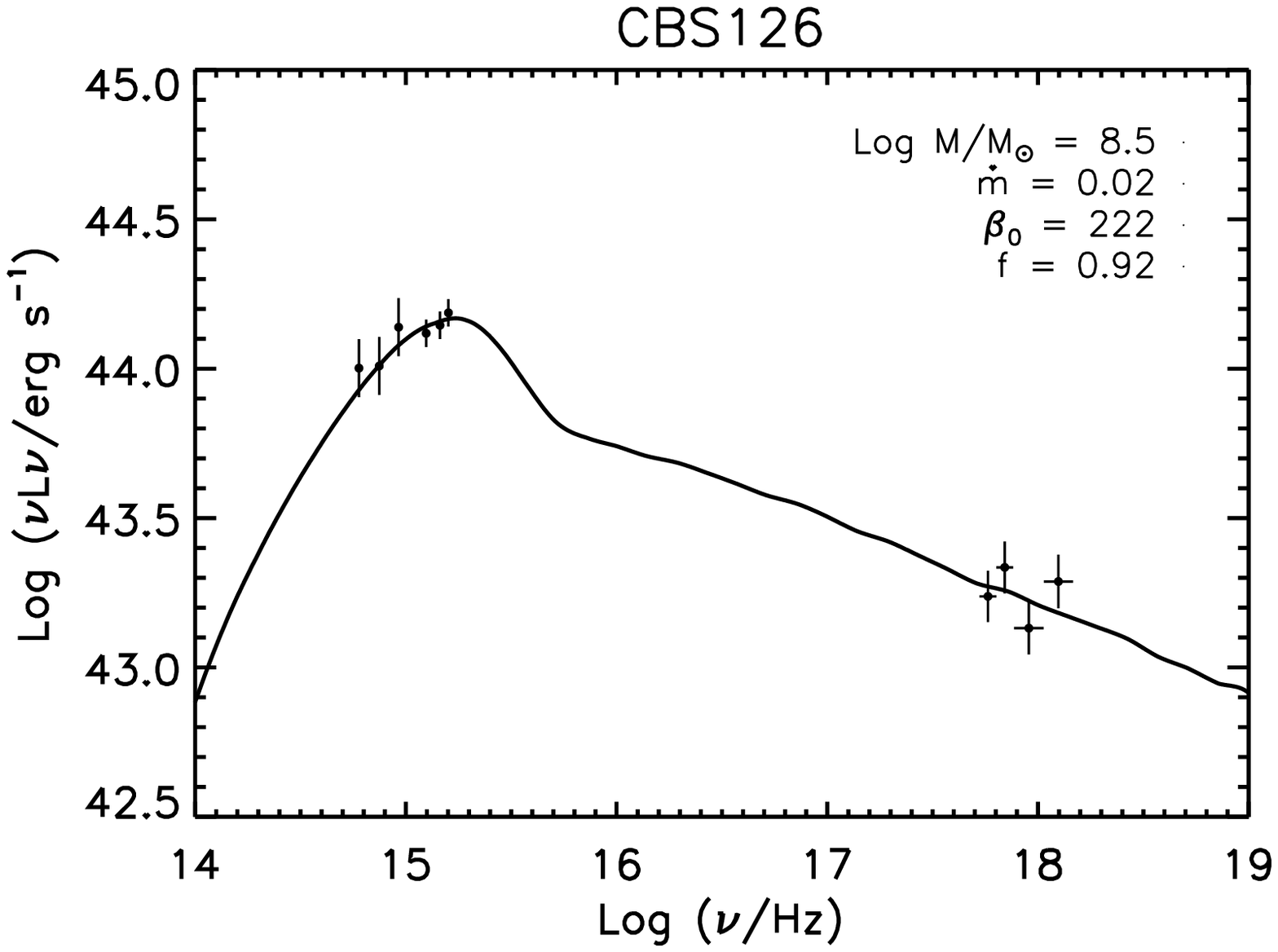}
	\includegraphics[width=5.7cm]{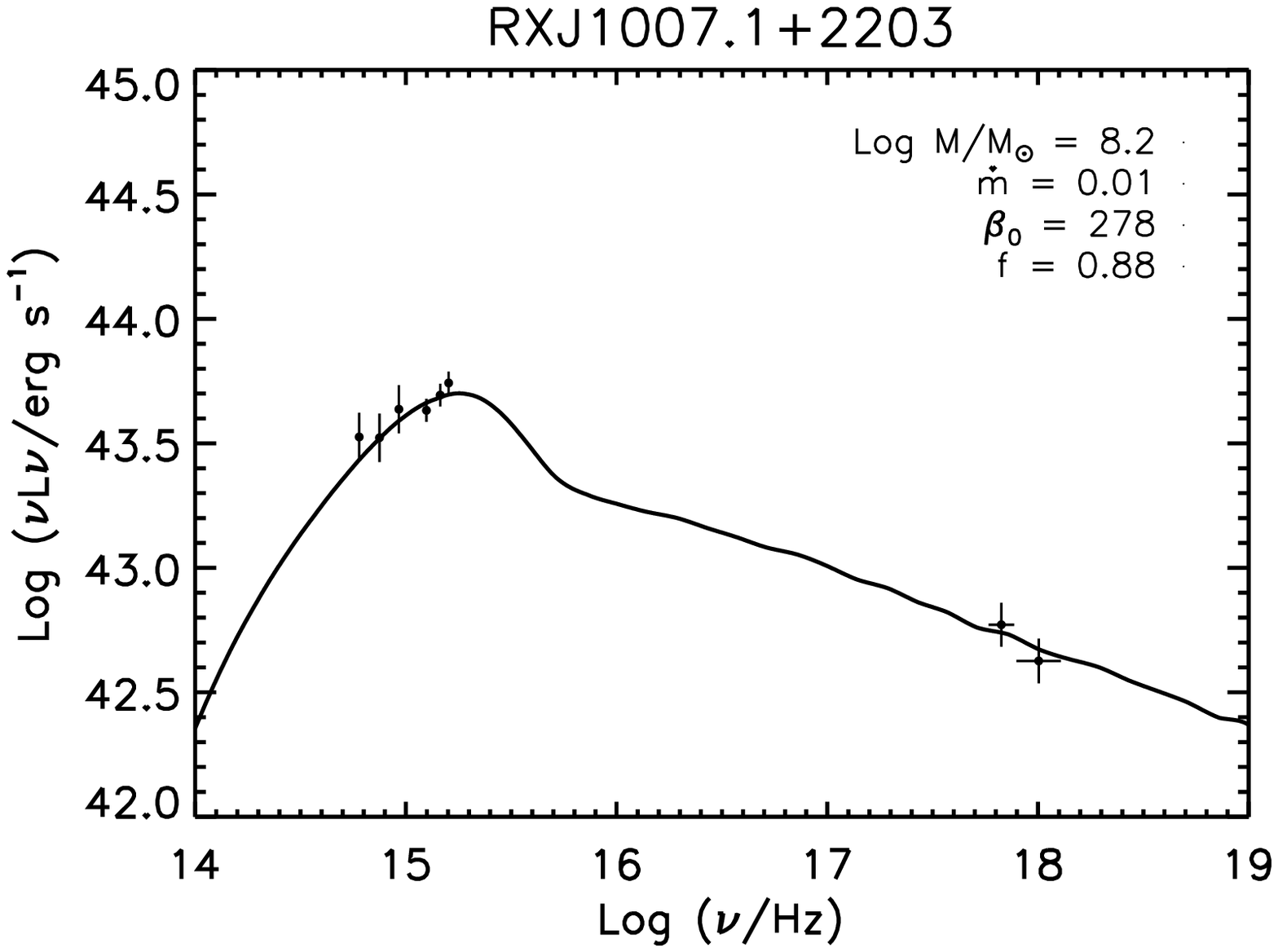}
	\includegraphics[width=5.7cm]{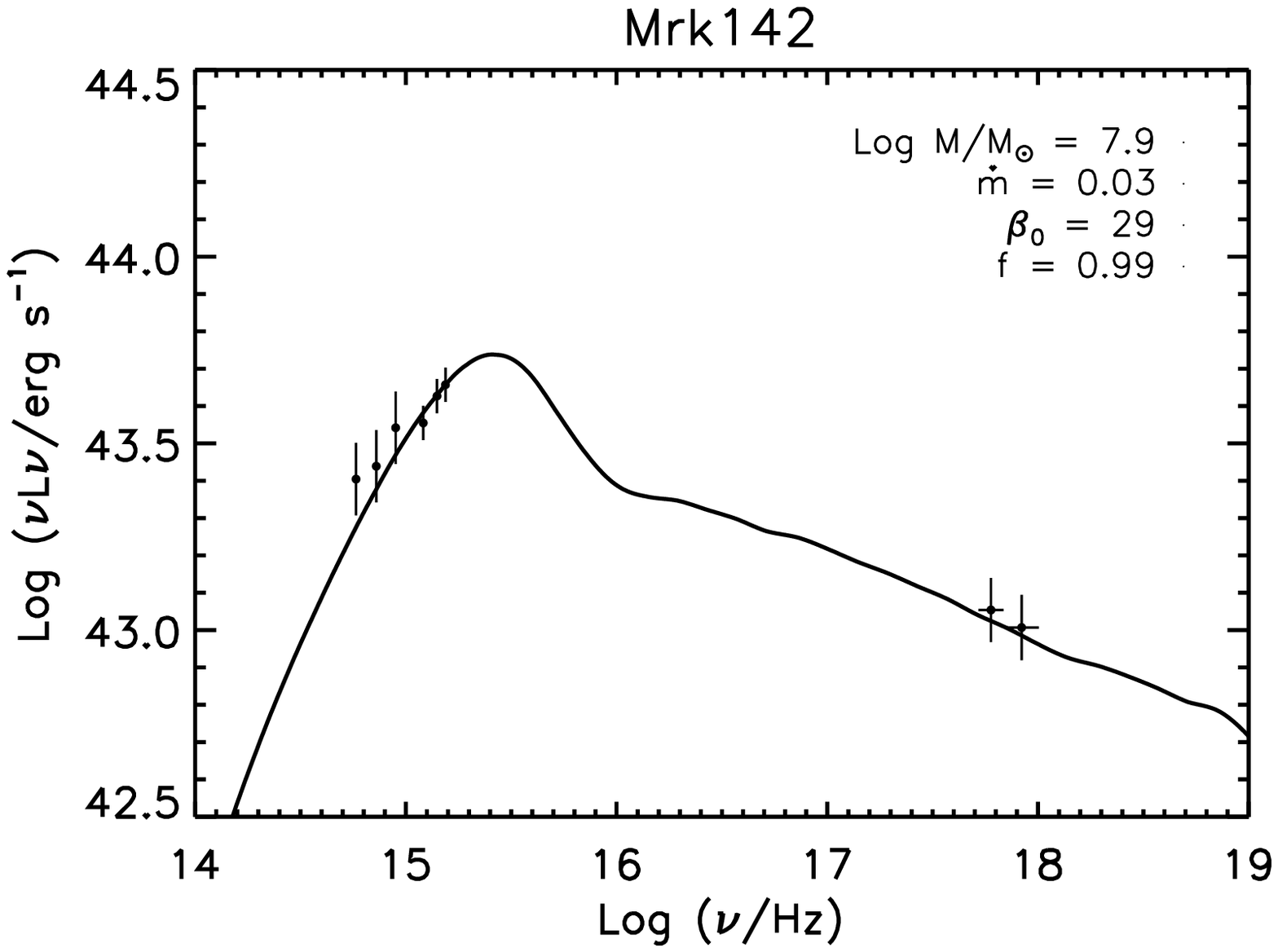}
	\includegraphics[width=5.7cm]{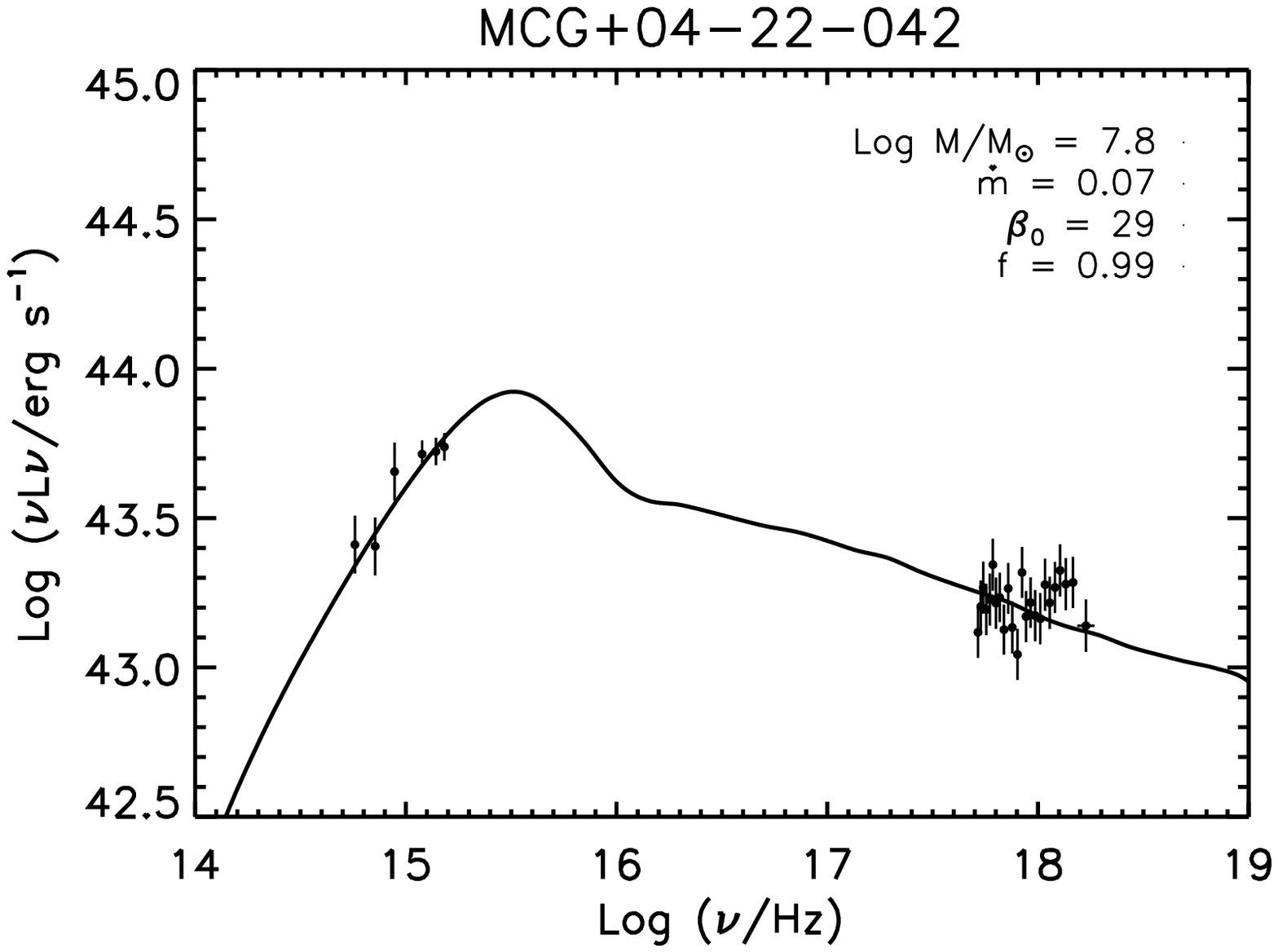}
	\includegraphics[width=5.7cm]{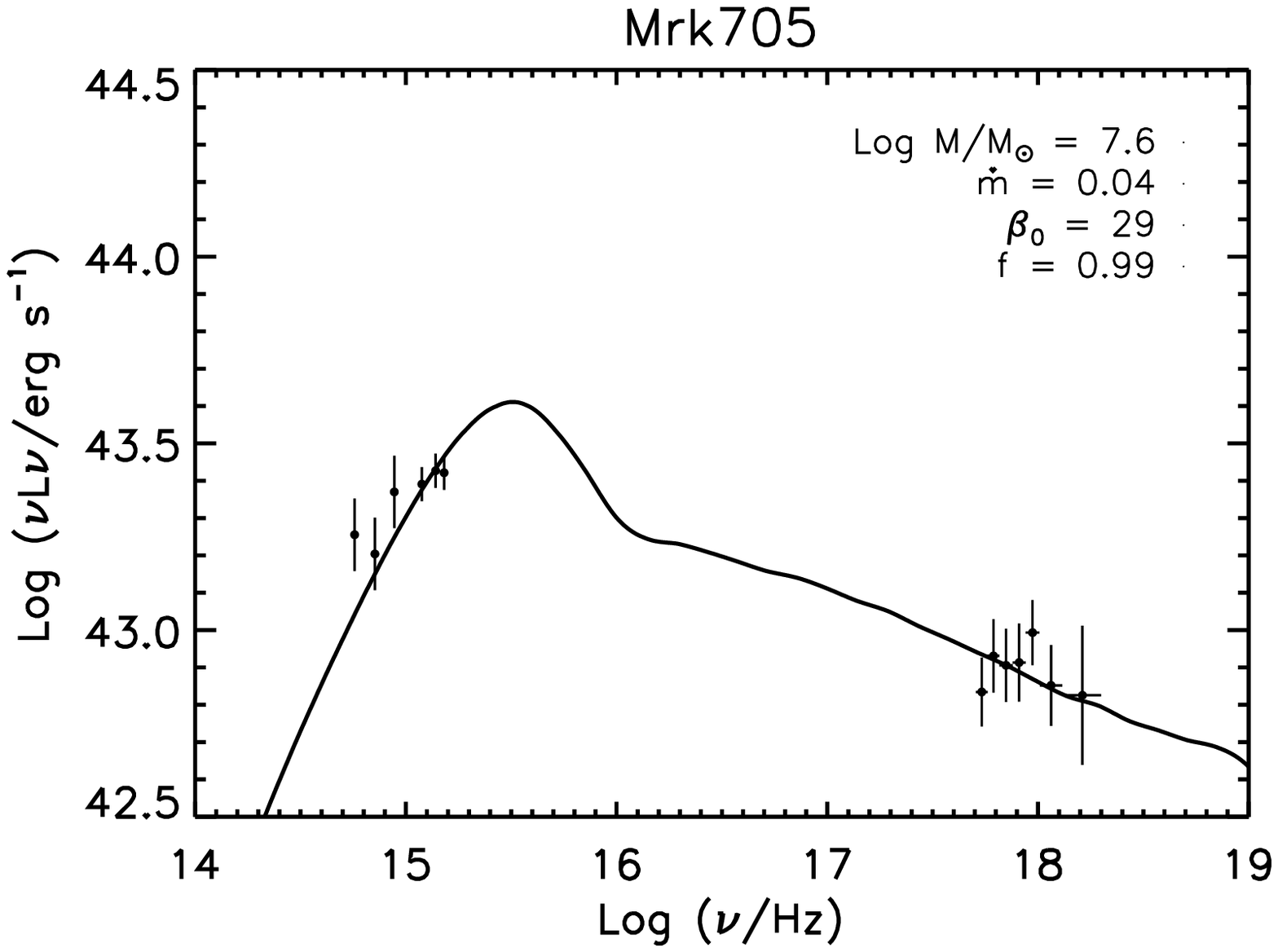}	
	\includegraphics[width=5.7cm]{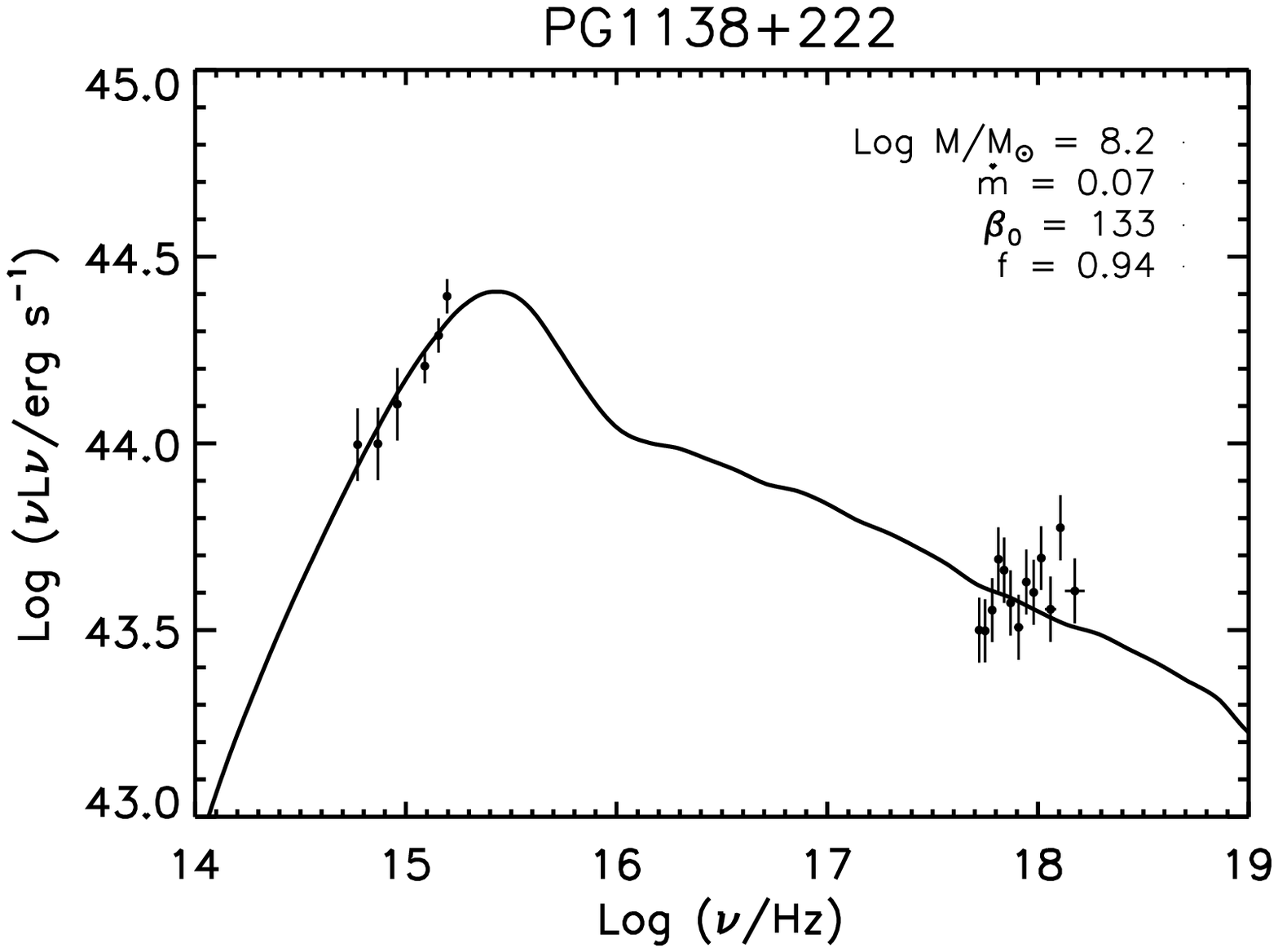}
	\includegraphics[width=5.7cm]{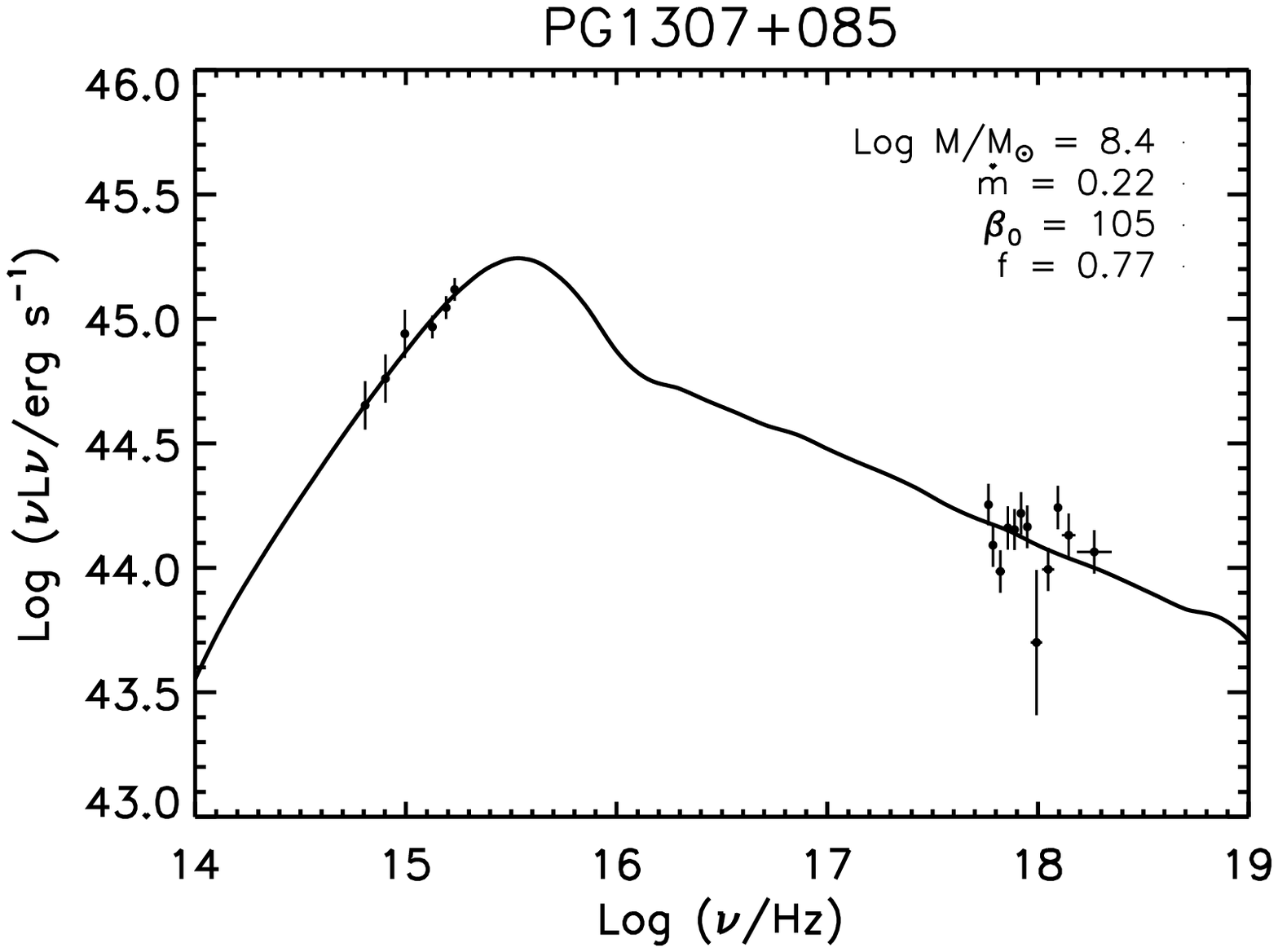}
	\includegraphics[width=5.7cm]{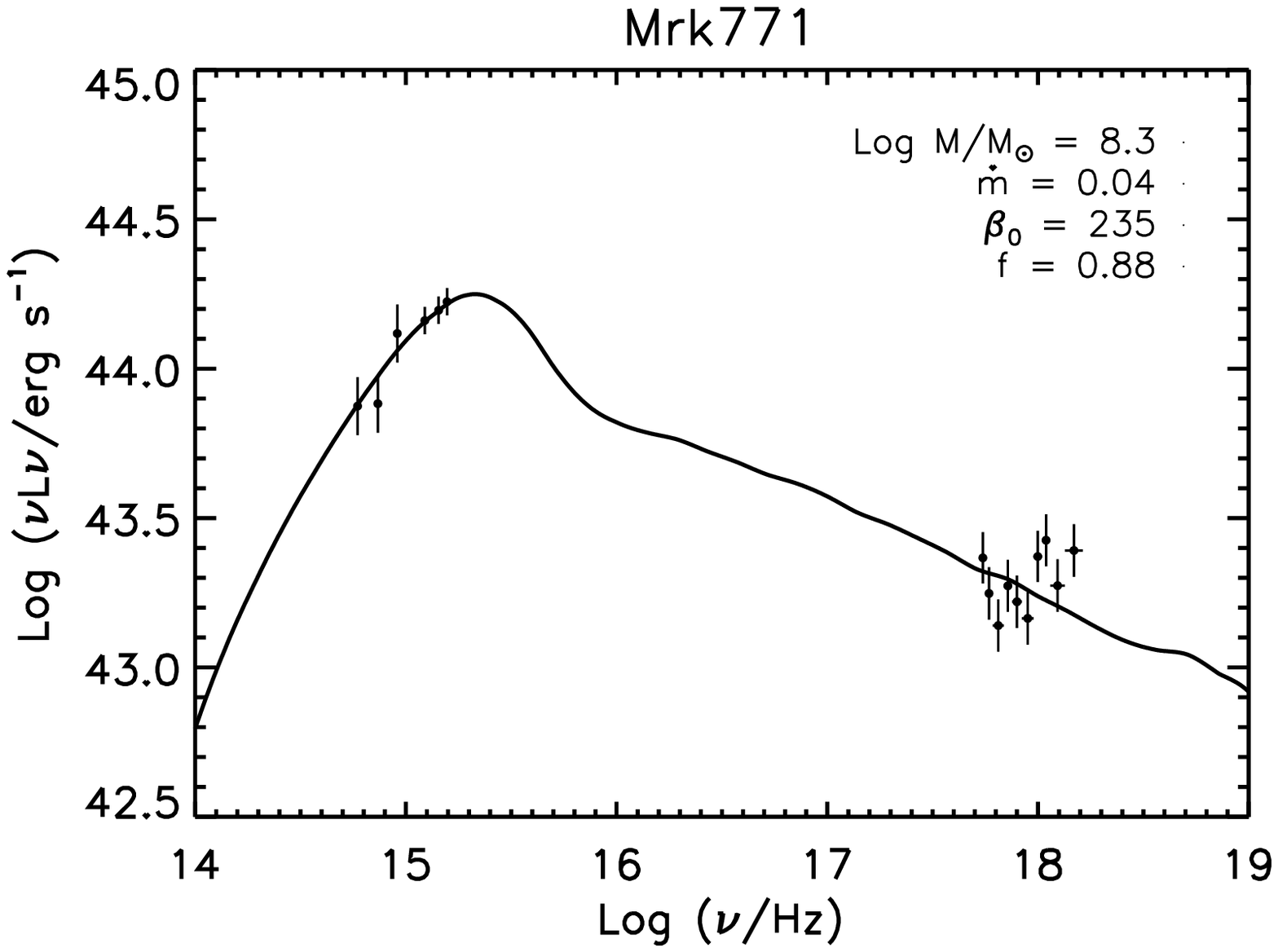}
	\includegraphics[width=5.7cm]{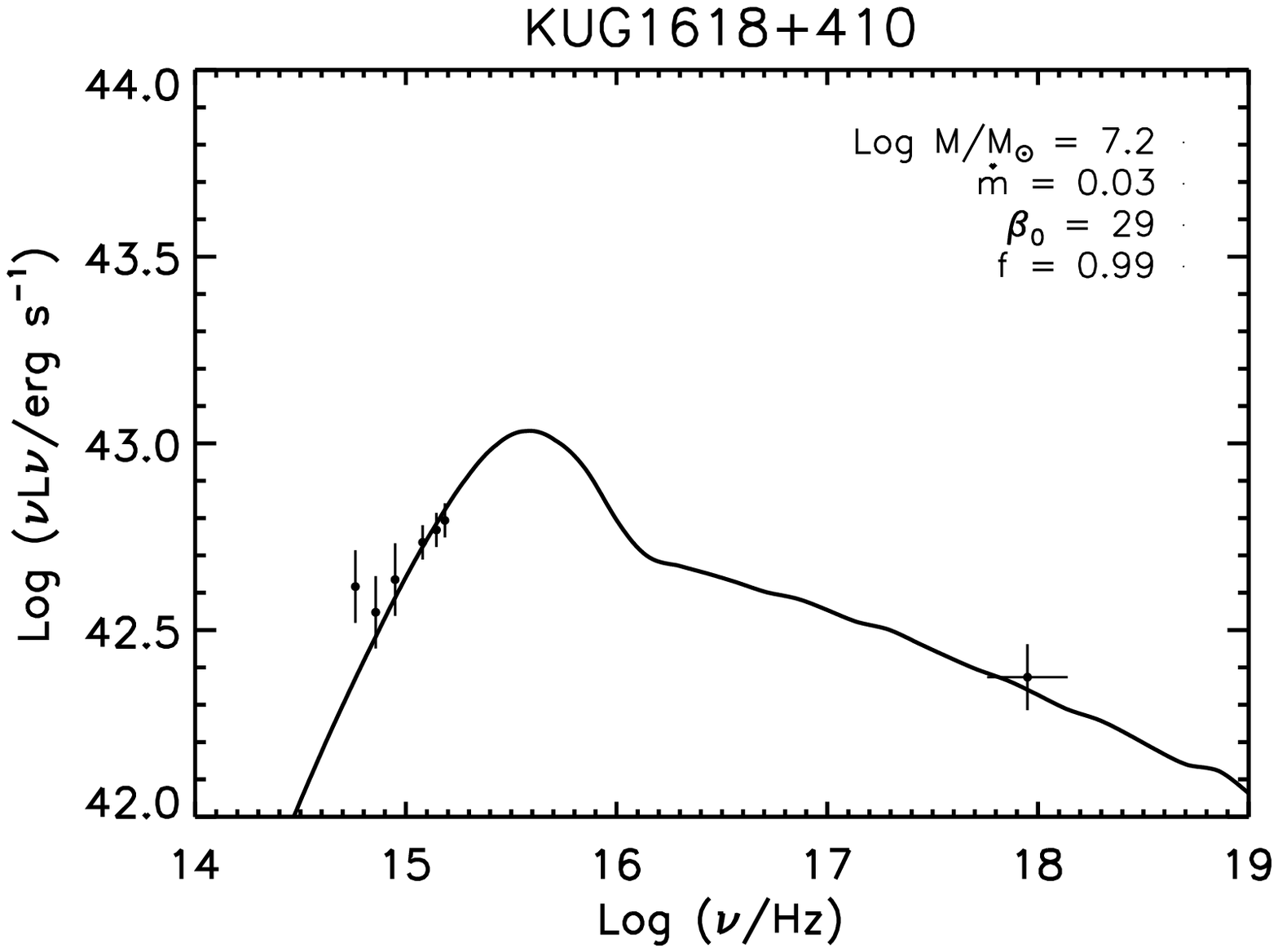}		
	\includegraphics[width=5.7cm]{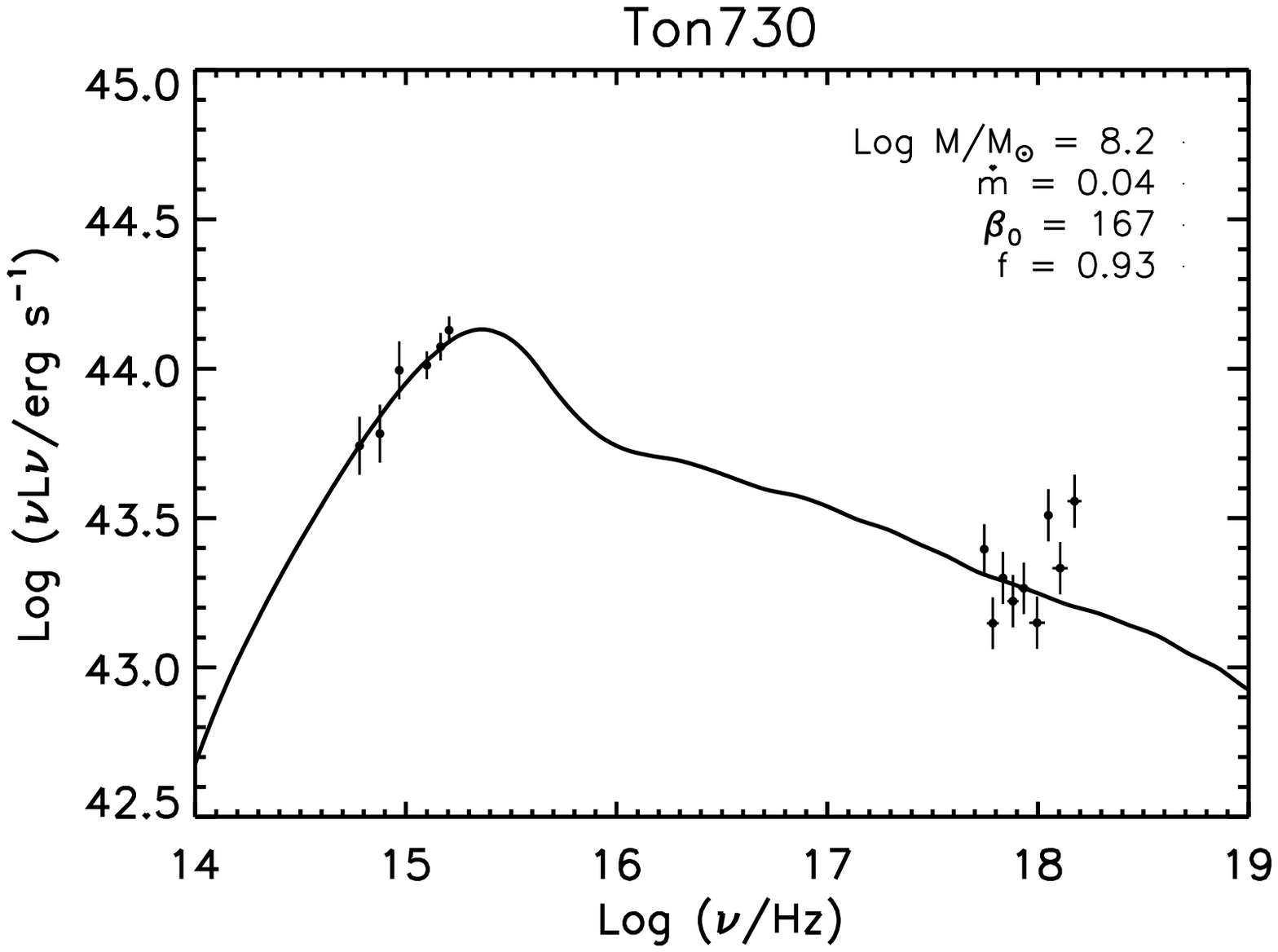}
	\includegraphics[width=5.7cm]{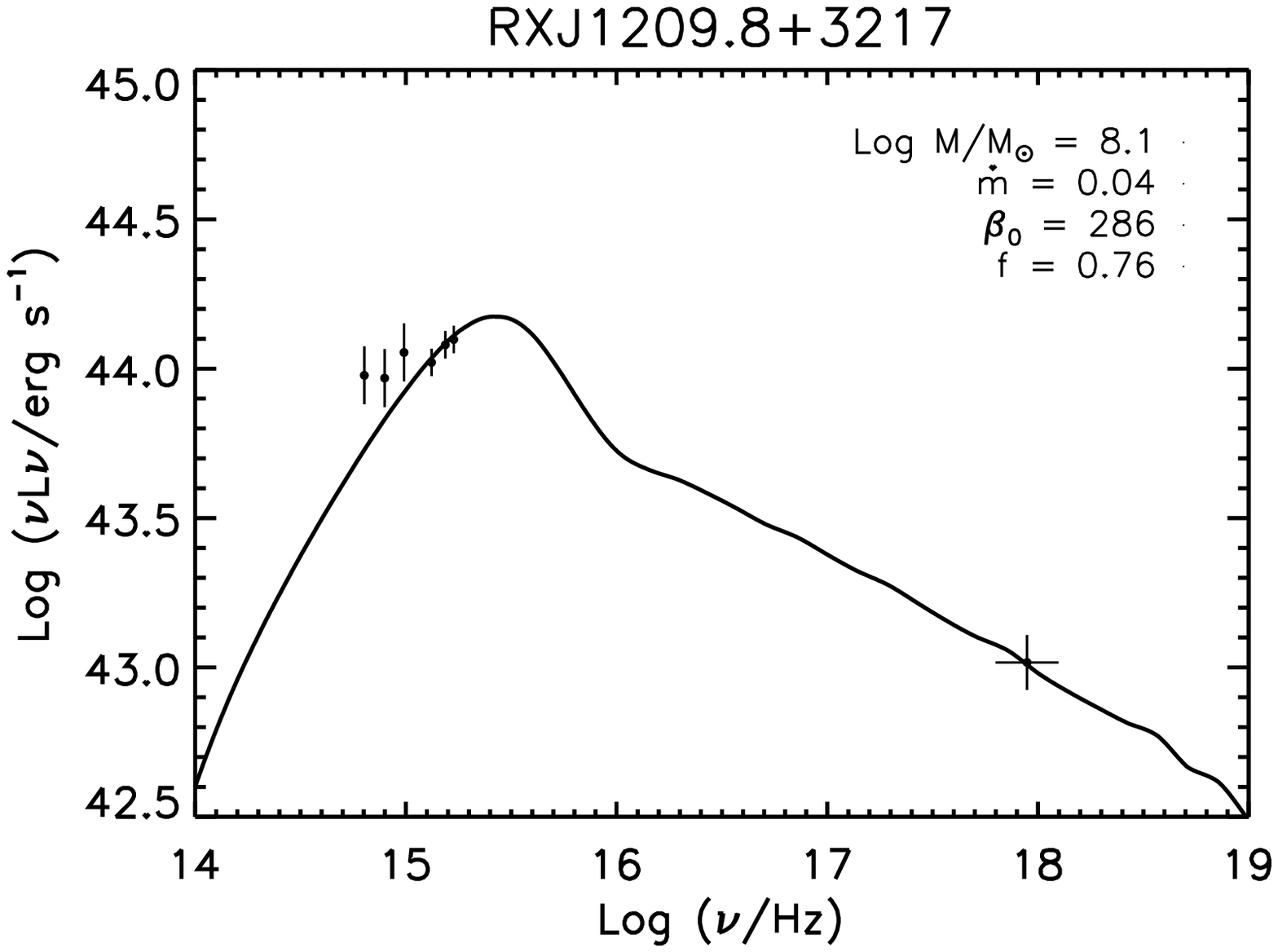}
	\caption{
	Broad-band SED data of the sample objects 
	in optical/UV and X-ray observed 
	simultaneously with {\it Swift} UVOT and XRT 
	\citep[taken from][]{2019MNRAS.487.3884C}, 
	 overplotted with the best-fitting SEDs of the
	 magnetic-reconnection-heated disc--corona model (solid line).
	The best-fitting parameters
	(black hole mass, accretion rate, and magnetic equipartition coefficient) 
	are labelled, as well as the derived fraction of energy
	released in the corona ($f$).
	}
	\label{fig:broadseds}	
\end{figure*}

\begin{figure*}
	\includegraphics[width=5.7cm]{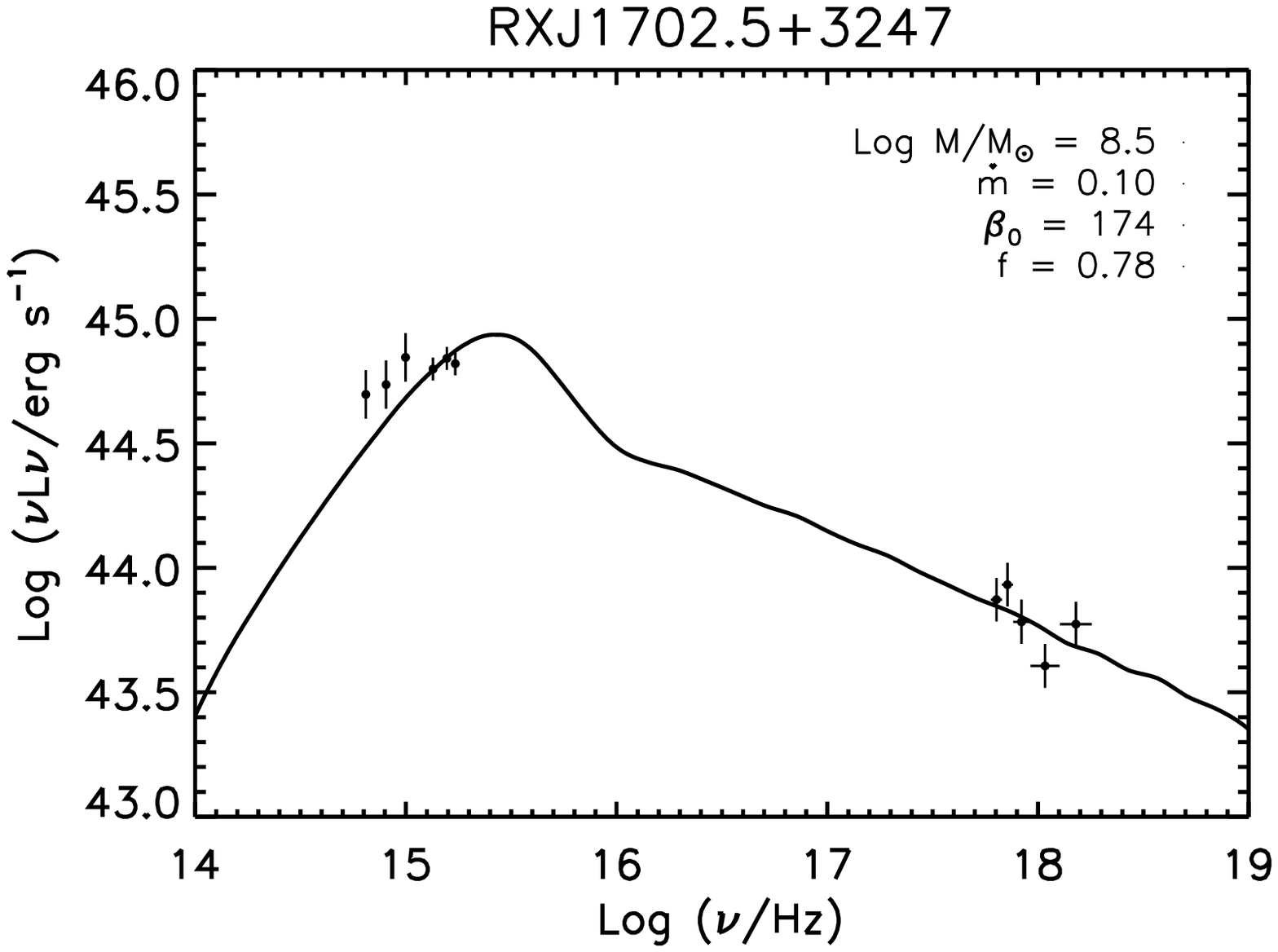}		
	\includegraphics[width=5.7cm]{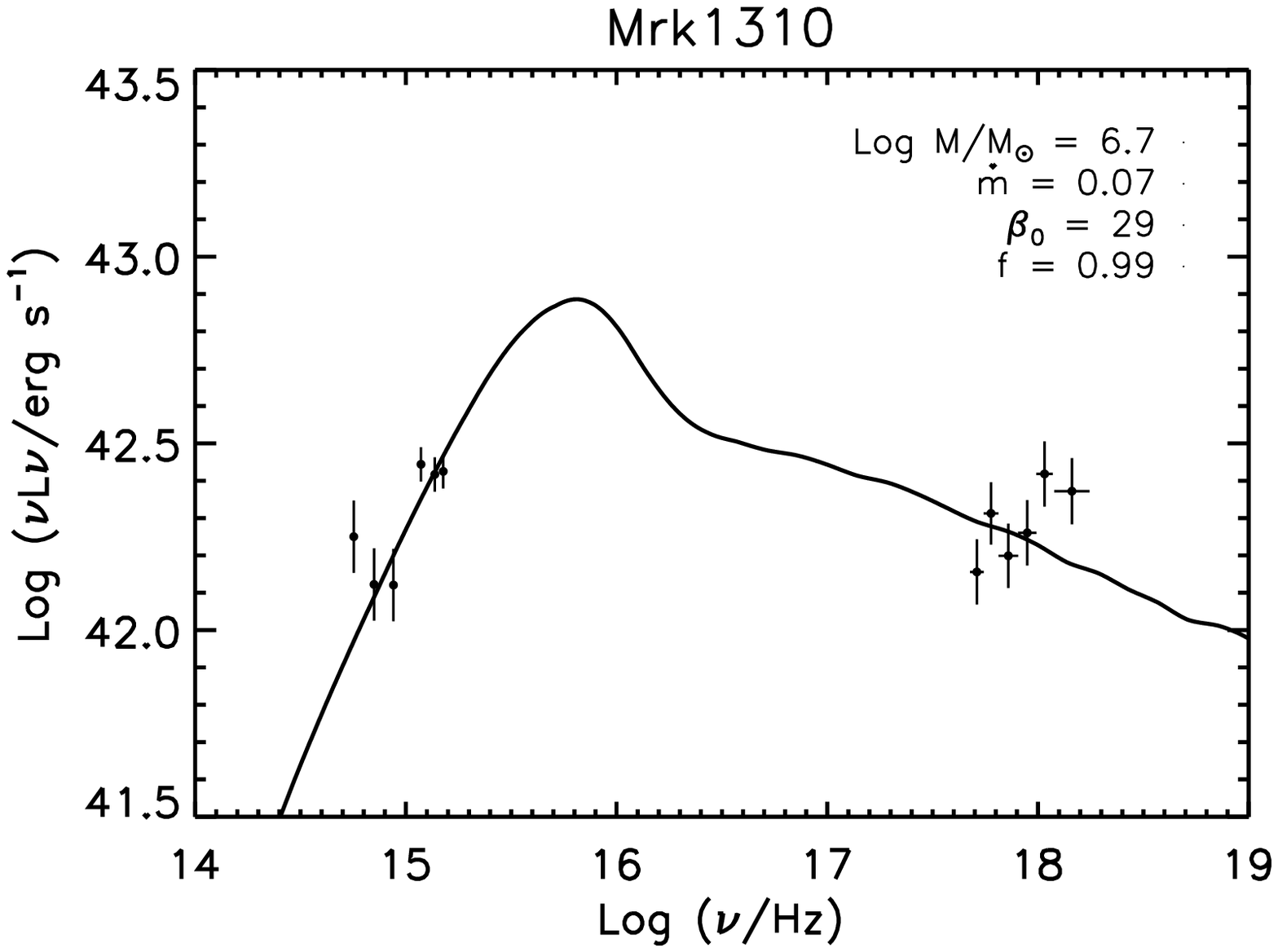}
	\includegraphics[width=5.7cm]{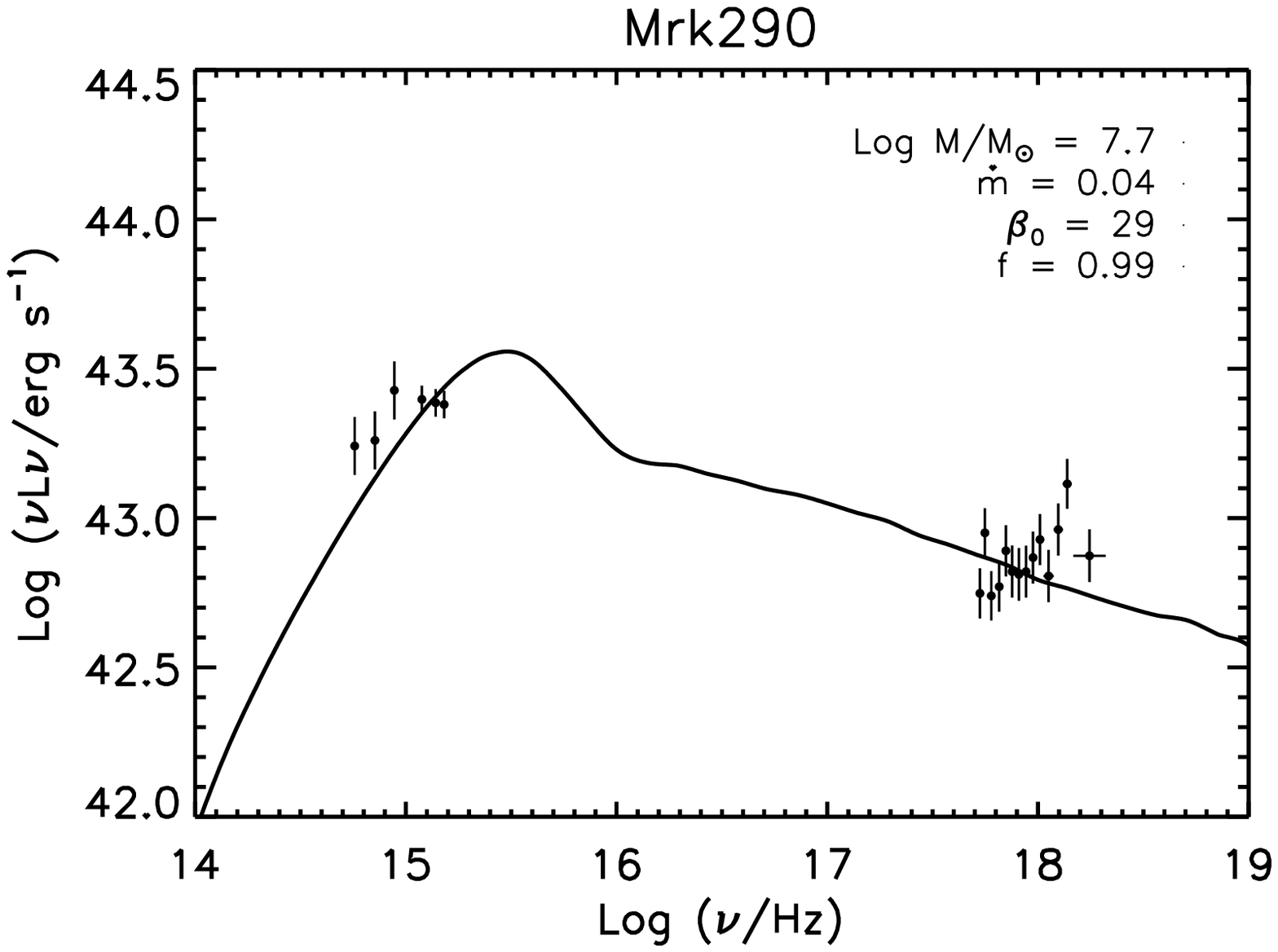}	
	\includegraphics[width=5.7cm]{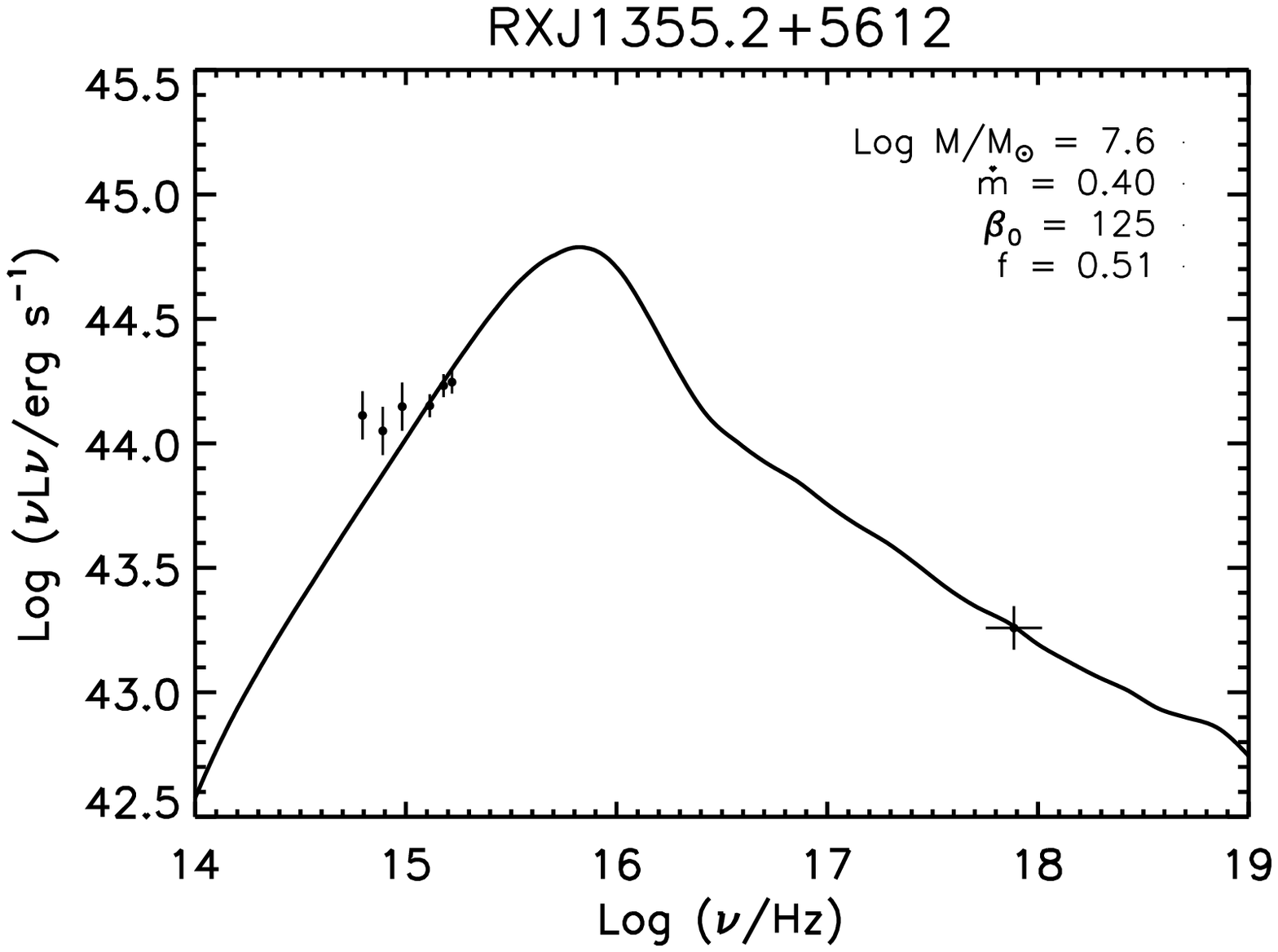}
	\includegraphics[width=5.7cm]{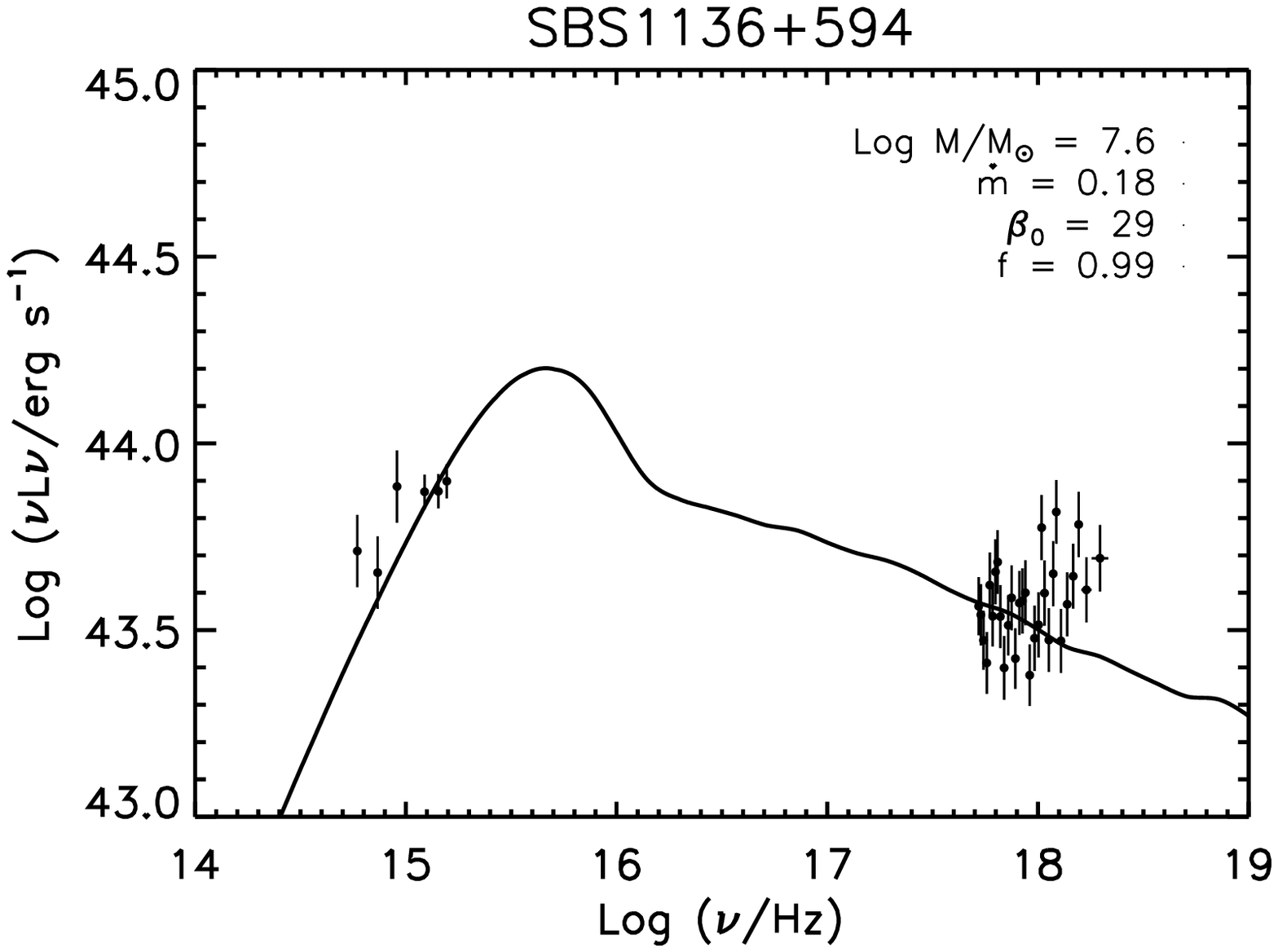}
	\begin{center}
	Figure~\ref{fig:broadseds} (continued)
	\end{center}
\end{figure*}

\begin{table*}
\centering
\caption{Fitting results and the virial black hole masses of sample objects}
\label{tab:params}
\begin{threeparttable}
\small
\renewcommand\arraystretch{1.3}
\begin{tabular}{ccccccccc}
\hline
\hline
Object & $\log (M_{\rm vir}/{\rm M_{\sun})}$ & $\log (M/{\rm M_{\sun}})$ & $\dot{m}$ & $\beta_0$ & ${\rm \chi^2_{\nu}}$ &  $p$ & $f$ \\
(1) & (2) & (3) & (4) & (5) & (6) & (7) & (8) &  \\
\hline
\hline
Mrk 1018 & $8.4$ & $7.8_{-0.5}^{+0.2}$ & $0.12^{+0.08}_{-0.04}$ & $95^{+13}$ & $7/16$ & $0.97$ & $0.96$ \\
Mrk 493 & $7.5$ & $8.1_{-0.2}^{+0.1}$ & $0.01^{+0.01}$ & $250_{-96}^{+57}$ & $5/11$ & $0.93$ & $0.89$\\
Mrk 1392 & $8.2$ & $8.0_{-0.2}^{+0.2}$ & $0.03_{-0.02}^{+0.02}$ & $111^{+99}$ & $10/17$ &$0.90$ & $0.96$ \\
Ton 1388 & $8.8$ & $8.6_{-0.2}^{+0.3}$ & $0.57_{-0.29}^{+0.43}$ & $111_{-24}^{+46}$ & $6/11$ &$0.87$ & $0.47$ \\
CBS 126 & $7.9$ & $8.5_{-0.3}^{+0.3}$ & $0.02_{-0.01}^{+0.02}$ & $222^{+63}$ & $4/7$ & $0.78$ & $0.92$ \\
RX J1007.1+2203 & $7.5$ & $8.2_{-0.4}^{+0.1}$ & $0.01^{+0.03}$ & $278_{-126}^{+98}$ & $3/5$ & $0.70$ & $0.88$ \\
Mrk 142 & $7.2$ & $7.9_{-0.3}^{+0.2}$ & $0.03_{-0.02}^{+0.06}$ & $29^{+167}$ & $3/5$ & $0.70$ & $0.99$\\
MCG 04-22-042 & $7.7$ & $7.8_{-0.2}^{+0.1}$ & $0.07_{-0.02}^{+0.08}$ & $29^{+98}$ & $24/27$ & $0.63$ & $0.99$ \\
Mrk 705  & $7.2$ & $7.6_{-0.2}^{+0.2}$ & $0.04_{-0.02}^{+0.06}$ & $29^{+114}$ & $9/10$ & $0.53$ & $0.99$ \\
PG 1138+222 & $7.6$ & $8.2_{-0.3}^{+0.2}$ & $0.07_{-0.03}^{+0.09}$ & $133^{+16}$ & $17/16$ & $0.39$ & $0.94$ \\
PG 1307+085 & $8.6$ & $8.4_{-0.3}^{+0.4}$ & $0.22_{-0.14}^{+0.48}$ & $105_{-35}^{+48}$ & $16/15$ &$0.38$ & $0.77$ \\
Mrk 771 & $7.9$ & $8.3_{-0.3}^{+0.2}$ & $0.04_{-0.02}^{+0.04}$ & $235_{-41}^{+26}$ & $16/13$ & $0.25$ & $0.88$\\
KUG 1618+410 & $7.1$ & $7.2_{-0.4}^{+0.4}$ & $0.03_{-0.02}^{+0.09}$ & $29^{+178}$ & $6/4$ & $0.20$ & $0.99$ \\
Ton 730 & $8.0$ & $8.2_{-0.2}^{+0.2}$ & $0.04_{-0.02}^{+0.03}$ & $167^{+28}$ & $20/12$ & $0.07$ & $0.93$ \\
RX J1209.8+3217 & $7.6$ & $8.1~{\rm (b)}$ & $0.04_{-0.01}^{+0.03}$ & $286_{-71}^{+52}$ & $9/4$ & $0.06$ & $0.76$ \\
RX J1702.5+3247 & $8.0$ & $8.5~{\rm (b)}$ & $0.10_{-0.02}^{+0.03}$ & $174_{-31}^{+18}$ & $15/8$ & $0.06$ & $0.78$\\
Mrk 1310 & $7.2$ & $6.7_{-0.2}^{+0.1}$ & $0.07_{-0.02}^{+0.06}$ & $29^{+102}$ & $20/9$ & $0.02$ & $0.99$ \\
Mrk 290 & $8.0$ & $7.7_{-0.2}^{+0.1}$ & $0.04_{-0.02}^{+0.02}$ & $29^{+130}$ & $31/17$ & $0.02$ & $0.99$ \\
RX J1355.2+5612 & $7.1$ & $7.6~{\rm(b)}$ & $0.40_{-0.06}^{+0.60}$ & $125_{-39}^{+29}$ & $15/4$ & $0.005$ & $0.51$ \\
SBS 1136+594 & $7.9$ & $7.6_{-0.2}^{+0.1}$ & $0.18_{-0.04}^{+0.14}$ & $29^{+45}$ & $59/33$ & $0.004$ & $0.99$ \\
\hline
\end{tabular}
{Notes}: 
Column (1): NED name of sample objects; 
column (2): virial black hole masses obtained from \citet{2019MNRAS.487.3884C};
column (3): black hole masses; 
those reaching the boundary of the allowed range  
($\log M_{\rm vir}\pm0.5$) are marked with 'b';
column (4): accretion rate; 
column (5): magnetic equipartition coefficient; 
column (6): reduced $\chi^2_\nu$; 
column (7): null hypothesis probability; 
and column (8): averaged fraction of energy released in corona. \\
Uncertainties quoted are at the 90 per cent level for one interesting parameter;
no uncertainties given mean that the parameters cannot be constrained. 
\end{threeparttable}
\end{table*}

The best-fitting parameters of the disc--corona model are listed in Table \ref{tab:params}
and the fitted SEDs are overplotted in Fig. \ref{fig:broadseds}.
As can be seen, the models can generally reproduce the overall SEDs across 
the optical/UV and X-ray bands reasonably well. 
For the majority of the sample, the model fits are acceptable 
(or marginally) in the statistical sense ($p\ge0.1$ in 13), 
although in several objects  
the X-ray spectral quality is too poor to yield meaningful constraint. 
In the remaining objects, the fits are poor
as the model apparently fails to account for the spectral slopes in 
either X-ray (four objects) or optical/UV (three objects).
For the overall sample, the modelled 2--10\,keV X-ray photon indices 
are in the range of $\Gamma=2.2$--$2.5$
(without taking into account the disc-reflection component). 
These values, however, appear to be steeper than those of many Seyferts 
which have $\Gamma \sim 1.7$--$1.9$ as directly observed and 
$1.9$--$2.0$ for the primary continuum when
the reflection component is accounted for
\citep[e.g.][]{1994MNRAS.268..405N, 2011A&A...530A..42C, 2013ApJ...772..114R, 2016MNRAS.458.2454L, 
2017ApJS..233...17R}, 
although AGNs with relatively higher Eddington ratios 
tend to show steeper $\Gamma$ \citep[e.g.][]{2006ApJ...646L..29S, 2008ApJ...682...81S, 2013MNRAS.433.2485B}.
See Section \ref{sec:slabcorona} for further discussion.  

\begin{figure}
\centering
	\includegraphics[width=\columnwidth]{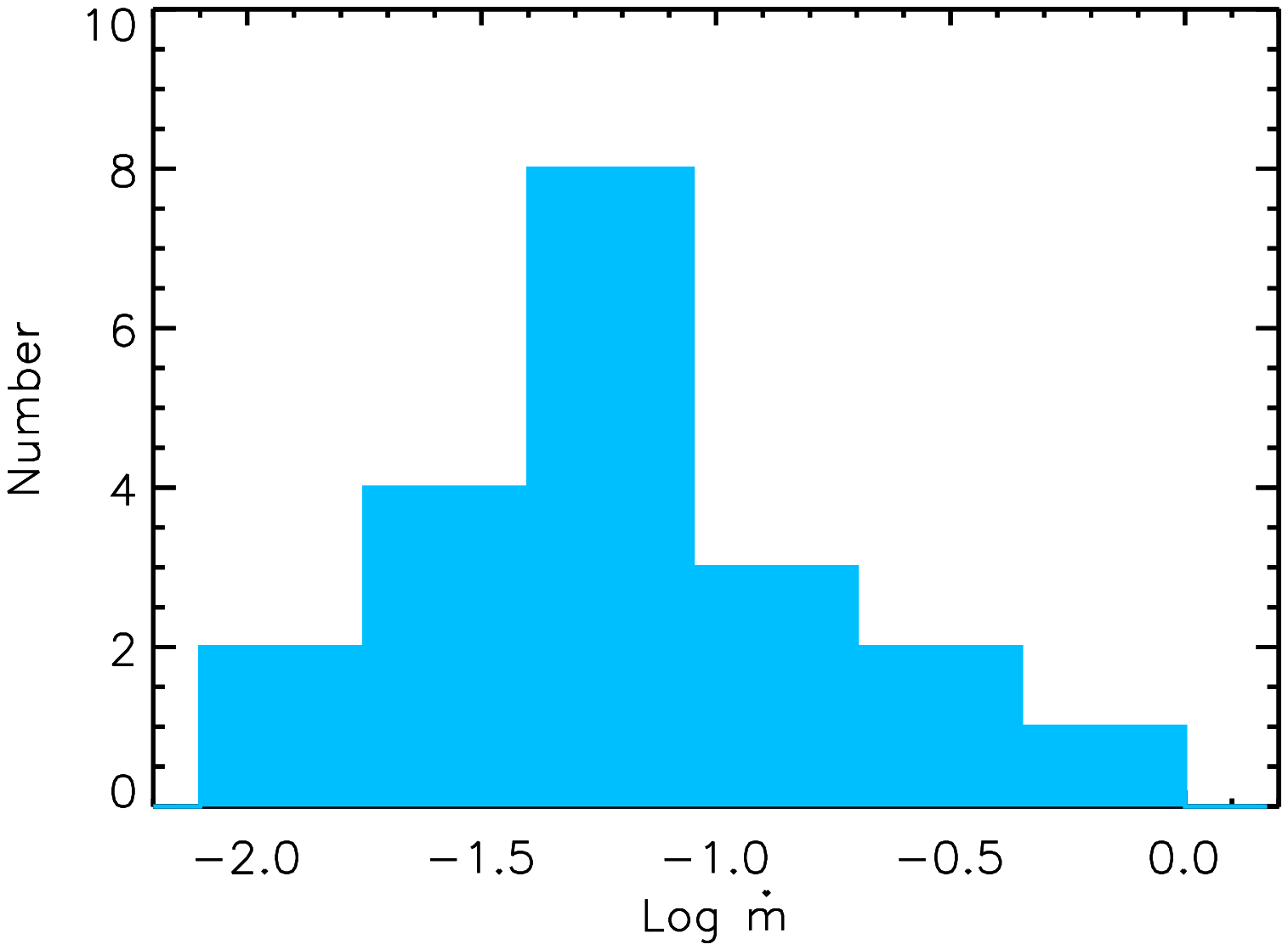}
	\caption{Distribution of the best-fitting mass accretion rates $\dot{m}$ of the sample objects. } 
	\label{fig:mdot}
\end{figure}

The distribution of the fitted mass accretion rates 
is plotted in Fig. \ref{fig:mdot}. 
It shows that the accretion rates centre around $0.04$, 
with a minimum $0.01$ and a maximum $0.57$. 
In particular, $\dot{m}$ is smaller than $0.3$ in $18$ out of the 20 objects. 
In only two sources relatively larger $\dot{m}$ are yielded, 
namely Ton 1388 ($0.57$) and RX J1355.2+5612 ($0.4$). 
Therefore, the distribution of $\dot{m}$ from the spectral fitting supports the basic assumption of our model, that the gravitational energy is efficiently released via 
the standard Shakura--Sunyaev disc ($\eta\approx0.1$).

\begin{figure}
\centering
	\includegraphics[width=\columnwidth]{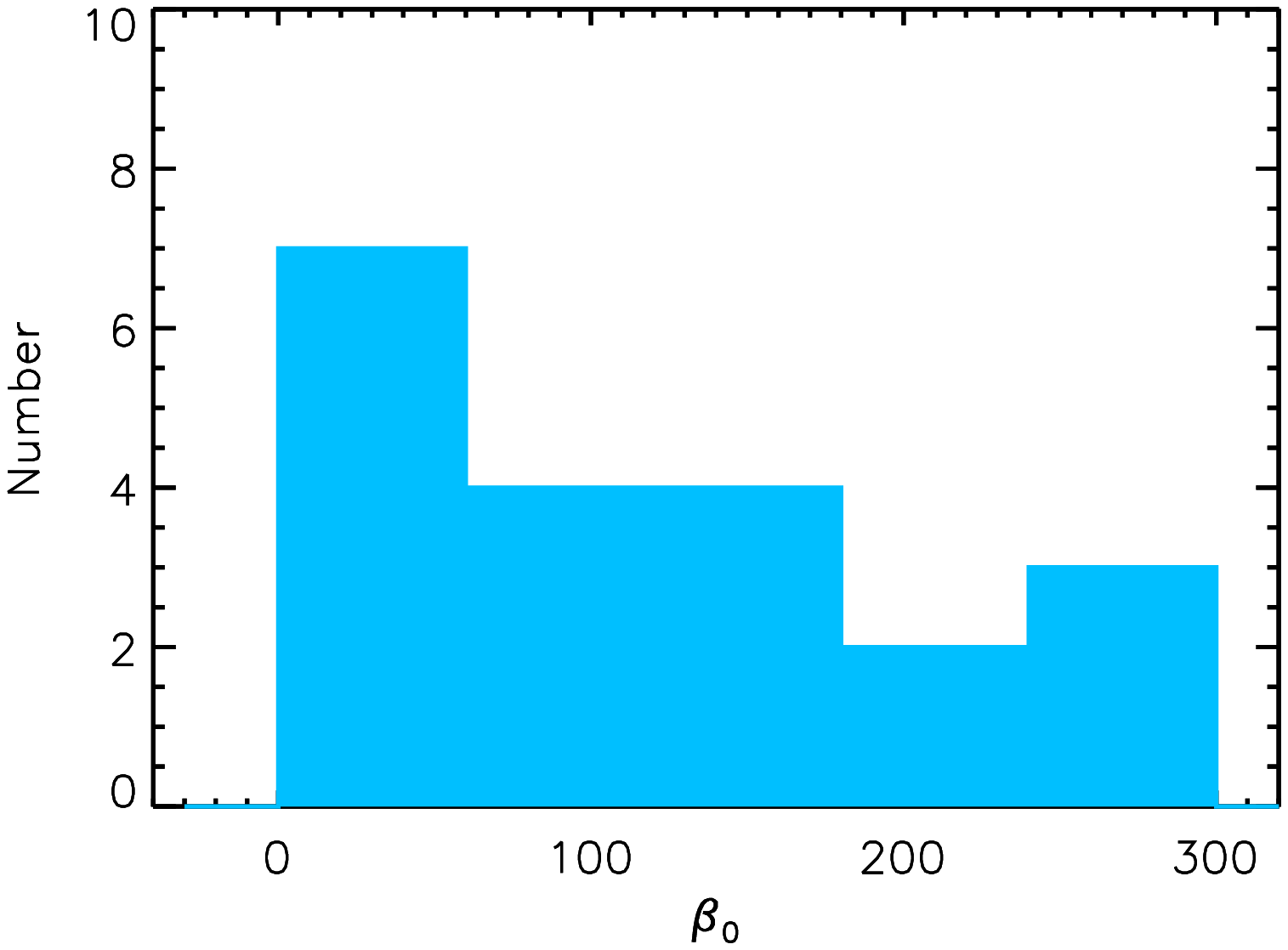}
	\caption{Distribution of the best-fitting magnetic equipartition coefficients $\beta_0$ of the sample objects.}
	\label{fig:beta}
\end{figure}

The distribution of $\beta_0$ is plotted in Fig. \ref{fig:beta}, showing a range from $\sim30$ to $\sim300$ with a median of $111$. 
The relative small values of $\beta_0$ suggest a strong magnetic-reconnection-heated corona residing in most of the sources. 
The energy fraction released in the corona (averaged along radius) 
exceeds 
80 (90) per cent in 15 (12) objects, 
as listed in Table \ref{tab:params}. 
In such cases, 
the intrinsic disc emission is so weak that the reprocessed radiation 
from the downward Compton scattering completely dominates the energy density of the soft photons. 
This implies that the radiation pressure in the disc (mid-plane) is much weaker than that in a standard disc for the same accretion rate. 
Specifically, from the spectral fitting we find that 
the innermost radius of the disc region where the radiation pressure
is indeed less than or comparable to the gas pressure 
can generally extend down to 
$R \le\ \sim20\,R_{\rm S}$ for $f \ga 0.9$
and all the way down to the last stable orbit for $f$ close to $1$ (the exact radius depends also on $\dot{m}$). 
An important consequence is that the radiation pressure-induced disc instability is ruled out in these AGN. This explains why there is no observational evidence for the instability which is predicted to exist in a standard disc around supermassive black hole. The magnetic field plays a key role in both producing strong X-ray emission and stabilizing the accretion flows in AGN. 

\begin{figure}
\centering
	\includegraphics[width=\columnwidth]{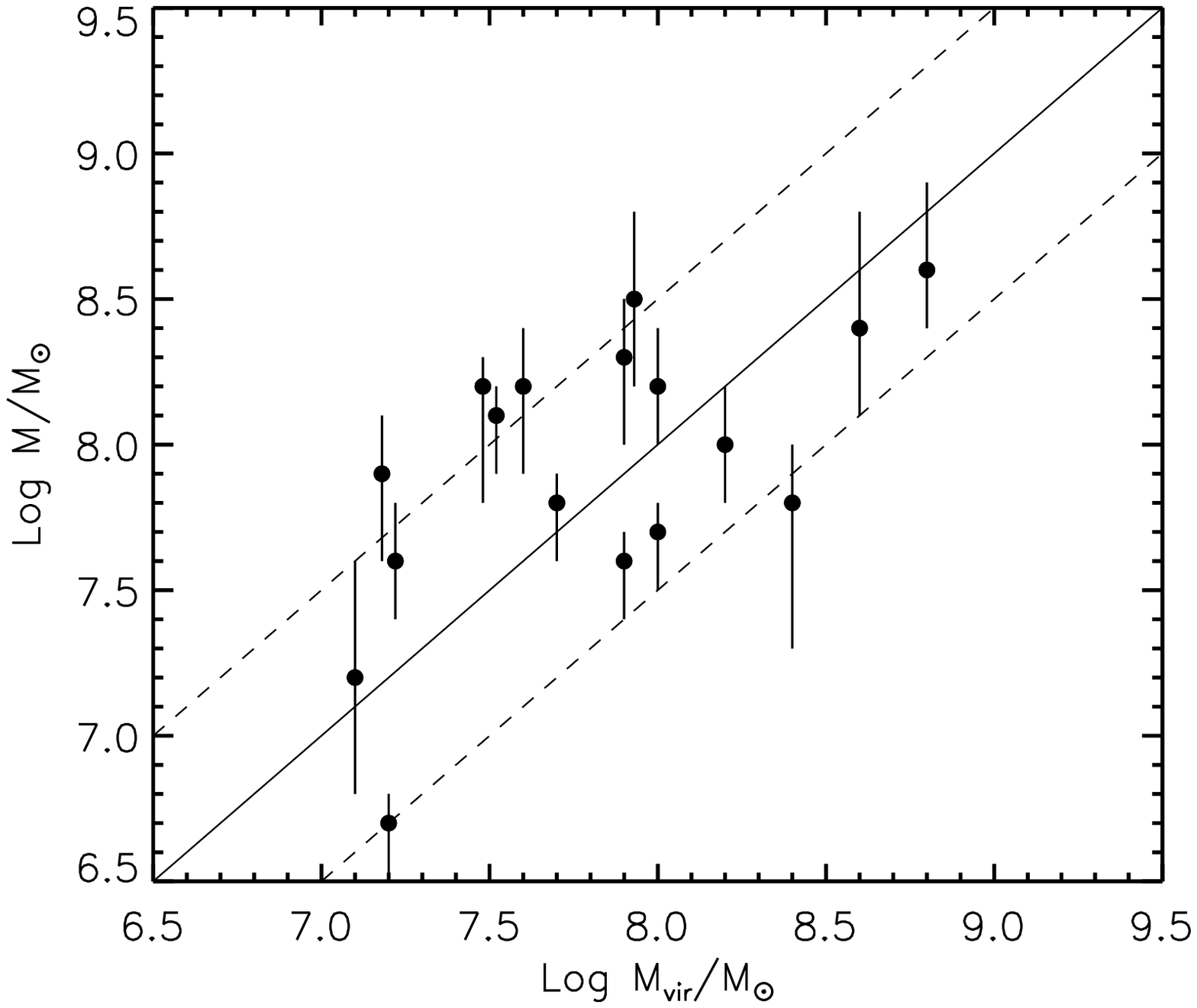}
	\caption{Comparison of the 
	fitted black hole masses and the virial masses
	\citep[from][]{2019MNRAS.487.3884C}. 
	Only 17 objects with fitted mass values consistent with $M_{\rm vir}$  
	within the mutual uncertainties are plotted;
	whereas for the others, the mass values reach the end of the allowed range 
	($\log M_{\rm vir} \pm 0.5$\,dex). 
	The error bars are at the 90 per cent confidence level.  
	The solid line represents the one-to-one relation, and
	the dashed lines denote the uncertainties ($\pm 0.5$\,dex) 
	of the virial mass. }
	\label{fig:bhmass}
\end{figure}

For the remaining five objects, weaker X-ray emission 
(relative to the optical/UV component) is found 
and a weaker corona solution is deduced. 
Less energy is carried away by the magnetic field, 
leading to a decrease in the coronal energy fraction $f$. 
Among these, three 
(RX J1209.8+3217, PG 1307+085, and RX J1702.5+3247)
have $f=0.7$--$0.8$ and thus 20--30 per cent
of the accretion energy is released in the disc;
the radiation pressure-dominant region begins to appear in the inner disc.
For the other two objects,  
which have the highest accretion rates among the sample,
Ton 1388 
($\dot{m}\approx0.57$ and $L_{\rm bol}\approx2.5 \times10^{46}~ {\rm erg ~ s^{-1}}$) 
and RX J1355.2+5612 
($\dot{m}\approx 0.4$ and $L_{\rm bol}\approx 1.6\times10^{45}~ {\rm erg ~ s^{-1}}$),
about half of the accreted energy ($f\sim 0.5$) is released in the disc.  
The inner regions of their discs (up to radii $R <\ \sim100\,R_{\rm S}$)
are completely dominated by radiation pressure,
although the reduced heating and hence the disc radiation can mitigate somewhat the photon-trapping effect therein. 
Compared to other AGNs in the sample, 
a strong disc and relatively weak corona scenario is favoured for them.

The black hole mass can, in principle, also be determined
from the spectral modeling. 
It would be interesting to compare the thus inferred masses
with those derived independently via the virial method 
from the single-epoch SDSS optical spectra
in our previous work \citep{2019MNRAS.487.3884C}, 
as given in Table \ref{tab:params}. 
For the majority ($17$) of the sample objects, 
the fitted mass values are 
broadly consistent with $M_{\rm vir}$ 
within their mutual uncertainties 
(0.5\,dex for $M_{\rm vir}$ and the $90$ per cent error for the fitted mass).
A comparison of the fitted masses with $M_{\rm vir}$ for these 17 objects 
is shown in Fig.\ref{fig:bhmass}.
Whereas for the remaining three objects, 
the mass value just reaches the end of the allowed range
(reaching the higher end $\log M_{\rm vir} +0.5$ for all).
We note that this bias of fitting the mass in these objects
arises mainly from  matching their relatively flat/red optical/UV slopes observed
with the turnover of the disc emission bump, 
which requires a higher black hole mass so as to shift 
the peak to a lower frequency
(see Section \ref{sec:slabcorona} for discussion).

\section{discussion}
\label{sec:discuss}

\subsection{Spectral deviations from disc--corona model prediction}
\label{sec:slabcorona}

In several objects
(particularly SBS 1136+594, Mrk 1310, Mrk 290, and Ton 730),
the observed X-ray spectra are too flat/hard 
\citep[$\Gamma\approx1.4-1.7$;][]{2019MNRAS.487.3884C}
to be accounted for by the model. 
Whereas in several others (e.g.\ RX J1007.1+2203, Mrk 142, and KUG 1618+410) 
the X-ray spectral data are of too low quality to 
 constrain strongly the model.
For the overall sample, our model generally yields   
 $\Gamma=2.2$--$2.5$, 
which appear to be steeper than those often
observed in typical Seyfert galaxies
\citep[$\Gamma \sim 1.7$--$1.9$; e.g.][]{1994MNRAS.268..405N, 2007A&A...461.1209D, 2008A&A...485..417D, 2011A&A...530A..42C, 2016MNRAS.458.2454L}. 
Here we discuss this discrepancy from both observational and theoretical considerations.

The {\em observed} X-ray spectra of some AGN may show a complex shape and deviate significantly from the original form of the primary corona emission. 
First, any absorption in excess of that is accounted for by models (usually the Galactic absorption) or partial coverage of the emission region will harden the observed X-ray spectra. 
Secondly, the X-ray spectra of some AGN are observed to be significantly contributed, or even dominated, by a reflection component from the accretion disc and dusty torus, which is peaked around $30$--$40$ keV and can flatten the observed overall spectra below $10$ keV \citep[e.g.][]{1991MNRAS.249..352G, 2005MNRAS.358..211R}, 
and not included in our modeling.
Observationally it has been demonstrated that, 
in many Seyferts with typical $\Gamma=1.7$--$1.9$ as observed, their 
primary X-ray continua have actually steeper slopes of 
$\Gamma=1.9$--$2.0$ when the reflection component is accounted for
\citep[e.g.][]{1994MNRAS.268..405N, 2007A&A...461.1209D,2011A&A...530A..42C, 2013ApJ...772..114R, 2017ApJS..233...17R}.
This is also confirmed by our simulations that 
the observed X-ray spectra will flatten by 
$\Delta \Gamma=0.1$--$0.2$ 
when an assumed reflection component is included,
by using {\sc xspec} 
and its disc reflection model {\sc xillver} \citep[][]{2013ApJ...768..146G}. 
However, for the relatively steep X-ray slopes of our modeling, 
this effect is not sufficient to explain the rather 
flat spectral indices of $\Gamma \sim 1.7$--$1.9$, 
unless in some extreme conditions. 
For example,
this happens when part of the primary X-ray emission from corona is hampered to reach the observer by various processes, such as obscuration or light bending \citep[e.g.][]{2003MNRAS.340L..28F, 2004MNRAS.349.1435M}. 
The current X-ray data quality, as obtained with {\it Swift}/XRT, is insufficient to disentangle these complex components from the primary coronal emission. 
We defer more rigorous modeling of the AGN spectra by taking into account the reflection component to a future work.

Furthermore, it has recently been suggested that 
the primary X-ray spectral slopes 
of AGNs depend on the Eddington ratio $\lambda_{\rm Edd}$
\citep[e.g.][]{2013MNRAS.433.2485B, 2017MNRAS.470..800T, 2019MNRAS.487.2463W}. 
Specifically, AGNs showing flat spectra $\Gamma \le 2$
have generally lower Eddington ratios $\lambda_{\rm Edd} \le 0.1$, 
and the spectra steepen with the increase of  $\lambda_{\rm Edd}$. 
Since some of our sample objects
have $\lambda_{\rm Edd} > 0.1$, 
their X-ray spectral indices are expected to be $\Gamma > 2$, 
somewhat steeper than $1.9$--$2.0$ of the 
{\em primary} continuum as found in many Seyferts.
In fact, our model naturally predicts the trend of spectral steepening with 
increasing $\lambda_{\rm Edd}$.

However, we expect that the above factors may only alleviate somewhat, rather than eliminate, the discrepancy between the model and observed spectral indices, 
due to the limitation of this and other similar models. A theoretical limit to the hardness of the spectral slope can be understood as the following. In the disc--corona coupling model, around half of the coronal radiation is intercepted by the disc, which is mostly absorbed (or scattered) and eventually re-emits at lower energy bands. This implies that the disc emission, contributed by both the irradiation and intrinsic radiation, is usually stronger than that of corona. Only at the extreme case that all the viscous energy is transported into the corona ($f\approx1$), can the coronal luminosity be comparable with that of disc. With decreased disc emission the Compton cooling rate also decreases, leading to an increase in the coronal temperature and density. Therefore,  the Compton-$y$ parameter (defines as $y=\frac{4kT_{\rm e}}{m_{\rm e}c^2}\tau_{\rm es}\propto n_{c}T_{\rm e}$) reaches its upper limit at $f\approx1$, corresponding to the hardest possible X-ray spectral slope. In fact, this has been a long-standing problem for the non-truncated planar disc geometry around black holes \citep[for a detailed review, see][]{2018A&A...614A..79P}, and one way to resolve this issue is to reduce the the interception fraction of the coronal radiation. To reach such a situation, several models have been proposed beyond the canonical plane-parallel corona configuration, such as the patchy/clumpy/truncated disc--corona model \citep[e.g.][]{1979ApJ...229..318G, 1994ApJ...432L..95H, 1995MNRAS.272..291S} and the dynamically outflowing corona model \citep[e.g.][]{1999ApJ...510L.123B, 2001MNRAS.326..417M}. 

We now consider the spectral deviations in the optical/UV band in some of the objects, e.g. RX J1209.8+3217, RX J1355.2+5612, and RX J1702.5+3247. In fact, their observed spectra are redder than those of the other sample objects and than the prediction of the standard Shakura--Sunyaev disc \citep[][]{2019MNRAS.487.3884C}. This is most likely not caused by dust reddening since the sample objects were selected to be essentially free from dust extinction, as indicated by their Balmer decrement in the optical spectra \citep[][]{2019MNRAS.487.3884C}. As demonstrated by \citet{2019MNRAS.487.3884C}, the observed flattening of the optical/UV spectra is likely a consequence of the loss of accreted mass from outside inward, by means of outflows or disc/winds, hence leading to a flattened temperature profile from outside inward. Such an effect is not taken into account in the current disc--corona model. To summarize, additional processes or spectral components can practically complicate the appearance of the spectra as observed in AGN.

\subsection{Relative strength of the coronal X-ray radiation}
\label{sec:xrayweak}

For the majority of the sample objects, a strong corona with an irradiation-heated disc scenario ($f>{80}$ per cent) is suggested by our disc--corona model. In these objects, the X-ray emission is relatively strong with regard to optical/UV emission, arising from relatively strong magnetic fields. We suggest that this scenario is likely common in typical Seyfert galaxies. Indeed, our sample is biased to local X-ray bright AGN which are predominantly Seyfert galaxies with moderate accretion rates. As for high-luminosity quasars, the situation is less clear as there are only a few quasars in our sample. 
Specifically, the two objects with the highest accretion rates,
Ton 1388 \citep[$\dot{m}\approx0.57$, 
$L_{\rm bol}\approx 2.5\times10^{46}~ {\rm erg ~ s^{-1}}$;][]{
2017A&A...597A..79P} 
and RX J1355.2+5612 
($\dot{m}\approx 0.4$,
$L_{\rm bol}\approx 1.6\times10^{45}~ {\rm erg ~ s^{-1}}$),
 are of typical quasar luminosities.
Their observed SEDs require only about half of the accretion energy transported into the corona ($f\sim0.5$), mainly due to a much higher accretion rate than the others ($\dot{m}\sim0.57$ and $0.4$, respectively), and their observed X-ray luminosities are much lower than the optical/UV luminosities. 
We anticipate that this may be a commonplace in many luminous quasars with high accretion rates, whose SEDs are dominated by a strong optical/UV bump. However, this has to be tested in the future with a larger sample of quasars. We suggest that in AGNs there may exist a wide range of the relative magnetic field strengths, 
and hence of the fractions of energy transported into the corona, 
leading to the diverse relative strengths of the corona X-ray emission with respect to the optical/UV emission.

A small fraction of AGNs are known to have significantly weak intrinsic X-ray emission compared to their optical/UV emission, referred to as X-ray weak AGN \citep[e.g.][]{1998A&A...330..108Y, 2000ApJ...528..637B, 2004ApJ...610..737W, 2012ApJ...761...73D, 2014ApJ...794...70L}. They are characterized by much larger effective optical-to-X-ray spectral indices \citep[defined as $\alpha_{\rm ox}=-0.384\log L({\rm 2keV})/L(2500~\angstrom)$,][]{1979ApJ...234L...9T} than those expected from the $\alpha_{\rm ox}$--$L_{2500}$ relation ($L_{2500}$ is the monochromatic luminosity at $2500$ \angstrom). In the framework of our model, these AGNs can be understood if their magnetic fields in the disc are much weaker (larger $\beta_0$ values) than those in normal AGNs. In the extreme case when the coronal energy fraction is too low ($f\sim0$), the emergent spectrum will approach that of the standard Shakura--Sunyaev disc. As a simple comparison with observations, we calculate theoretical $\alpha_{\rm ox}$ for a given range of $L_{2500}=10^{27}$--$10^{28}$ ${\rm erg \ s^{-1} \ Hz^{-1}}$ assuming $m=10^6$ and $\beta_0={350}$. We find $\alpha_{\rm ox}$ spanning a range of $\sim1.5$--$1.8$, which are in good agreement with those of the X-ray weak AGNs in the \citet{2012ApJ...761...73D} sample. This suggests that the observed weak X-ray emissions in these objects can be attributed to a very weak magnetic activity of the underlying disc. 

\section{summary}
\label{sec:sum}

Aiming at the long-standing problem of how the disc corona is heated so as to sustain strong X-ray emission as observed in bright AGNs, 
we have studied a thin disc coupled with a corona through magnetic field and radiation in detail. 
The magnetic-reconnection-heating  model \citep{2002ApJ...572L.173L, 2003ApJ...587..571L, 2016ApJ...833...35L} has been refined
and applied to modelling the observed SED of a sample of AGNs. 
Our theoretical investigation shows that the model can  
produce a variety of broad-band SEDs 
for objects with various black hole masses, 
accretion rates and magnetic field strengths. 
The accretion rate and magnetic field play key roles in determining the X-ray spectral slope and the relative radiation strength between the disc and the corona. 
Specifically, a higher accretion rate leads to a steeper  X-ray spectrum and weaker X-ray emission (relative to the optical/UV emission), 
as a consequence of less evaporation;
a stronger magnetic field results in a flatter X-ray spectrum and stronger X-ray emission  as it transports more accretion energy from the disc to the corona;
the black hole mass, however, only shifts the peak wavelength of the optical/UV emission.
The magnetic field transports accretion energy into the corona, 
providing a solution to the problem of energy shortage in
corona as a strong X-ray emitter in AGNs. 
Meanwhile, this process also helps naturally eliminate or alleviate the unstable radiation pressure-dominant region in the central part of accretion discs in AGN.

The model is then applied to fitting the broad-band SEDs 
of a sample of $20$ local AGNs studied extensively 
in our previous work \citep{2019MNRAS.487.3884C}, 
which have simultaneous X-ray and optical/UV observations and well-measured AGN parameters. 
It is found that, in general, the overall observed broad-band spectral shapes 
can be reasonably reproduced by our disc--corona model,
although in some objects the X-ray spectra are too flat to be 
accounted for or too noisy to yield meaningful constraint.
In the majority of the sample (15/20),
the accretion energy is mostly transported into the corona ($f>80$ per cent)
by the magnetic field, 
supporting a strong corona coexisting with an irradiation-heated disc. 
Meanwhile, the fitting results rule out largely a
radiation pressure-dominant region in the central discs.
In only a few bright objects at quasar luminosities 
(e.g.\ Ton 1388 and RX J1355.2+5612), 
the fits lead to only about half of the energy released in the corona ($f \sim 0.5$), 
given their much higher observed optical/UV luminosities relative to the X-ray ones. 
The fitted black hole masses are generally consistent with the virial masses 
within their mutual uncertainties for the majority.

We note that our model generally predicts relatively steep 
2--10\,keV photon indices, $\Gamma=2.2$--$2.5$.
They appear to be steeper than 
the rather flat spectra observed in several sample objects,
as well as in many other Seyferts ($\Gamma < 2.0$),
though the discrepancy can be alleviated to some extent by 
considering some effects (e.g. contribution of a disc-reflection component).
Further theoretical investigation is needed to 
resolve this issue, which is generally inherent in  
models with similar disc--corona geometry as ours.

\section*{Acknowledgements}
The authors thank the referee for constructive suggestions that help to improve the paper. This work is supported by the National Program on Key Research and Development Project (grant no. 2016YFA0400804), the National Natural Science Foundation of China (grants  $11673026$, $11803047$, and $11773037$), the Strategic Priority Research Program of the Chinese Academy of Sciences (grant no. XDB23040100), and the Strategic Pioneer Program on Space Science, Chinese Academy of Sciences (grants  XDA15310300 and XDA15052100). HQ thanks Alice Breeveld, He-Yang Liu, and Bei You for useful comments and suggestions. Part of this work is based on the observations obtained by the Neil Gehrels {\it Swift} Observatory. We acknowledge the entire {\it Swift} team for providing the data that made this work possible. This research has made use of the NASA/IPAC Extragalactic Database (NED), which is funded by the National Aeronautics and Space Adminis- tration and operated by the California Institute of Technology.










\bsp	
\label{lastpage}
\end{document}